\documentclass[useAMS,usenatbib,psfig,eps,graphics]{mn2e}
\usepackage{natbib,psfig,graphicx,subfigure,epsfig,amsmath} 
 
\def\grtsim{\mathrel{\hbox{\rlap{\hbox{\lower2pt\hbox{$\sim$}}}\raise2pt\hbox{$>$}}}} 
\def\lesssim{\mathrel{\hbox{\rlap{\hbox{\lower2pt\hbox{$\sim$}}}\raise2pt\hbox{$<$}}}}

\def\degree{\nobreak\ifmmode{^\circ}\else{$^\circ$}\fi}

\newcommand{\kms}{km~s$^{-1}$}

\newcommand{\lbol}{$L_{\rm bol}$}

\newcommand{\lfir}{$L_{\rm FIR}$}

\def\lsol{L$_{\odot}$}

\def\msol{M$_{\odot}$}
\def\msolyr{M$_{\odot}$~yr$^{-1}$}

\def\bestfitT{12} 
\def\bestfitN{10$^{6.1}$}
\def\bestfitr{6.2}

\def\regAT{135} 
\def\regAN{10$^{3.8}$}
\def\regAr{0.7}

\def\margflatT{12} 
\def\margflatN{10$^{3.8}$}
\def\margflatr{0.7}

\def\margpriorT{90} 
\def\margpriorN{10$^{3.9}$}
\def\margpriorr{0.8}

\newcommand{\highCI}{[CI]$(^{3}P_{2}-^{3}P_{1})$}
\newcommand{\lowCI}{[CI]$(^{3}P_{1}-^{3}P_{0})$}

\newcommand{\aap}{A\&A}

\newcommand{\aj}{AJ}
\newcommand{\apj}{ApJ}
\newcommand{\apjl}{ApJL}
\newcommand{\apjs}{ApJS}

\newcommand{\araa}{ARA\&A}
\newcommand{\mnras}{MNRAS}
\newcommand{\nat}{Nat}

\newcommand{\myemail}{hana.schumacher@port.ac.uk}
 
\begin{document}

\title[Gas and dust in AMS12]{Gas and dust in a $z=2.8$ obscured quasar\thanks{Based on observations carried out with the IRAM Plateau de Bure Interferometer. IRAM is supported by INSU/CNRS (France), MPG (Germany) and IGN (Spain).}.}
\author[Schumacher et al.]
{Hana Schumacher$^{1,2}$\thanks{\myemail}, Alejo Mart\'\i nez-Sansigre$^{1,2}$, Mark Lacy$^{3}$, Steve Rawlings$^{4}$,\\
\newauthor Eva Schinnerer$^{5}$\bigskip \\
$^{1}$Institute of Cosmology and Gravitation, University of Portsmouth, Dennis Sciama Building, Burnaby Road, \\ Portsmouth, PO1 3FX, United Kingdom \\
$^{2}$SEP{\it net}, South-East Physics network\\
$^{3}$North American ALMA Science Center, National Radio Astronomy  Observatory, 520 Edgemont Road, \\ Charlottesville, VA 22903, United States of America\\
$^{4}$Astrophysics, Department of Physics, University of Oxford, Keble Road, Oxford, OX1 3RH, United Kingdom\\
$^{5}$Max-Planck-Institut f\"{u}r Astronomie, K\"{o}nigstuhl 17, D-69117 Heidelberg, Germany}

\date{}

\pagerange{\pageref{firstpage}--\pageref{lastpage}} \pubyear{}

\maketitle

\label{firstpage}
\vspace{-0.5 cm}

\begin{abstract}  
We present new detections of the CO(5-4), CO(7-6), \lowCI\ and \highCI\ molecular and atomic line transitions towards the unlensed, obscured quasar AMS12 ($z=2.7672$), observed with the Institut de Radioastronomie Millim\'{e}trique (IRAM) Plateau de Bure Interferometer (PdBI). This is the first unlensed, high redshift source to have both atomic carbon ([CI]) transitions detected. Continuum measurements between 70 $\mu$m and 3 mm are used to constrain the far infrared (FIR) spectral energy distribution (SED), and we find a best fit FIR luminosity of log$_{10}$[\lfir/\lsol] = $13.5\pm0.1$, dust temperature T$_{\rm{D}}$ = $88\pm8$ K and emissivity index $\beta = 0.6\pm0.1$. The highly-excited molecular gas probed by CO(3-2), (5-4) and (7-6), is modelled with large velocity gradient (LVG) models. The gas kinetic temperature T$_{\rm{G}}$, density n(H$_{2}$), and the characteristic size $r_0$, are determined using the dust temperature from the FIR SED as a prior for the gas temperature. The best fitting parameters are T$_{\rm{G}}$ = \margpriorT\ $\pm8$ K, n(H$_{2})$ = $10^{3.9\pm0.1}$ cm$^{-3}$ and $r_0 = 0.8\pm0.04$ kpc. The ratio of the [CI] lines gives a [CI] excitation temperature of $43\pm10$ K, indicating the [CI] and the high-excitation CO are not in thermal equilibrium. The [CI] excitation temperature is below that of the dust temperature and the gas kinetic temperature of the high-excitation CO, perhaps because [CI] lies at a larger radius where there may also be a large reservoir of CO at a cooler temperature, perhaps detectable through the CO(1-0). Using the \lowCI\ line we can estimate the strength of the CO(1-0) line and hence the gas mass. This suggests that a significant fraction ($\sim30\%$) of the molecular gas is missed from the high-excitation line analysis, giving a gas mass higher than that inferred from the assumption that the high-excitation gas is a good tracer of the low-excitation gas. The stellar mass was estimated from the mid-/near-infrared SED to be M$_{\star} \sim 3\times10^{11}$ \msol. The Eddington limited black hole mass is found from the bolometric luminosity to be M$_{\bullet}\grtsim1.5\times10^{9}$ \msol. These give a black hole - bulge mass ratio of $M_{\bullet}/M_{\star}\grtsim0.005$. This is in agreement with studies on the evolution of the $M_{\bullet}/M_{\star}$ relationship at high redshifts, which find a departure from the local value $\sim0.002$. We discuss the implications for the evolution of the black hole in AMS12 and its host galaxy.

\end{abstract}

 
\section{Introduction} \label{sec:intro}

Studies into the gas and dust in high redshift ($z\grtsim2$) quasar hosts allow us to observe an important epoch in galaxy formation. Determining the physical properties of the gas and dust allow for the characterisation of these galaxies, providing comparisons to the nearby Universe. 

The gas and dust are thought to be heated by nearby star formation (SF) (i.e. heating by young OB stars), and possibly by the AGN itself. Investigations into possible AGN heating of the gas and dust are important to distinguish the SF and AGN contributions to the far-infrared luminosity (\lfir), and the CO luminosity. 

The host galaxies of obscured quasars are easier to observe and characterise since the ultraviolet and optical emission from the central engines is obscured along the line of sight by intervening gas and dust. Some obscured quasars have been proposed to be at an evolutionary phase where the host galaxies contain more gas and dust \citep{Sanders1988,Fabian1999}. Therefore obtaining stellar, gas and dust mass estimates, inferring SFR and the properties of the quasar, can reveal the connections between the quasar and its host, and test the evolution of the ratio between the black hole mass ($M_{\bullet}$) and the host galaxy's bulge properties, i.e. stellar velocity dispersion ($\sigma_{\star}$), luminosity and mass ($M_{bulge}$). 

To derive physical properties of the molecular gas, the most dominant tracer of H$_{2}$, carbon monoxide (CO), is observed. Observing multiple CO rotational transition lines provides the CO spectral energy distribution (SED), or ``CO ladder". This can be used to infer the physical properties, i.e. the kinetic temperature and density, of the molecular gas via the fitting of detailed models such as large velocity gradient (LVG) models \citep{Scoville1974,Goldreich1974}. 

Other emission lines such as HCN, [CI] and [CII], have been observed right out to the highest redshift galaxies often with the help of gravitational lensing \citep[e.g][]{SolVan2005,Walter2011}. The atomic carbon molecule [CI] is thought to map the CO emission, with its critical density very close to that of CO(1-0) ($\sim10^{2}$ cm$^{-3}$). Also, studies of the [CI] and CO in Orion by \citet{Ikeda2002} show the emission comes from the same regions. [CI] can be described by the two optically-thin lines of a 3-level system and we can use the line ratio to directly determine the properties of the gas. 

The unlensed, obscured quasar,  AMS12 (R.A.(J2000) = 17 18 22.65, dec(J2000) = +59 01 54.3), at redshift $z~=~2.767$, has been observed at multiple wavelengths \citep{Martinez2005,Martinez2006a, Martinez2006b,Martinez2009,Klockner2009}. The redshift was determined from the Ly{\small$\alpha$}, C{\tiny{\textrm{IV}}} and He{\tiny{\textrm{II}}} lines in the optical spectrum \citep{Martinez2006a}. Its mid-IR SED suggests the galaxy corresponds to a progenitor of the present-day $\sim2$L$^{*}$ galaxies \citep{Martinez2005,Martinez2006a, Martinez2006b}. A strong detection of the CO(3-2) line was first presented in \citet{Martinez2009}, prompting further investigation into the CO ladder. 

In this paper we report on the detections of two higher rotational transitions of CO in AMS12, along with the detections of the \lowCI\ and \highCI\ lines. Section 2 details the observations. Section 3, is dedicated to the dust continuum and the fit to the far-infrared (FIR) SED of AMS12. Section 4 details further analysis into the molecular gas properties using LVG models. Atomic carbon in AMS12 is investigated further in Section 5, the results are discussed in Section 6. A summary of our findings is presented in Section 7. We assume a cosmology with $\rm{H_{0}}=70~\rm{kms^{-1}~Mpc^{-1}}$, $\Omega_{\Lambda} = 0.70$ and $\Omega_{\rm{m}}=0.30$ throughout this paper.

\section{Observations} 
\subsection{IRAM PdBI observations}\label{sec:gas_obs} 

The CO(3-2) ($\nu_{\rm{rest}} = 345.796$ GHz) transition, was observed in 2009 using IRAM's Plateau de Bure Interferometer (PdBI), in the $3~\rm{mm}$ band centred on $91.796~\rm{GHz}$ \citep{Martinez2009}. The CO(5-4) and CO(7-6) rotational transitions ($\nu_{\rm{rest}} = 576.268$ GHz and 806.652 GHz respectively) were observed with the PdBI 2 mm and 1.3 mm bands centred on 152.986 and 214.148 GHz for 4 nights during April and May 2010.

These observations used 6 antennas in the compact D configuration, with dual polarization utilising the narrow-band correlator. The \textit{WideX} correlator was used for the CO(5-4) and CO(7-6) observations and covers a bandwidth of $3.6 ~\rm{GHz}$ with a fixed spectral resolution of 2 MHz. The flux and phase calibrators for all observations were MWC349 and 1637+574 respectively. The bandpass calibrator for the 3 mm, 2 mm, and 1.3 mm bands was 3C345, and 3C273 was used as the bandpass calibrator for one night of the 1.3 mm observations. The observations were made with generally good weather conditions (precipitable water vapour (PWV) ranging from 2 -8 mm), however, the 1.3 mm band was observed over three nights due to bad weather on those nights (generally conditions worsened at the end of the observing runs with the PWV reaching $>10$ mm on some days).  

\begin{figure*}
 \centering
\subfigure[CO(3-2)]{
\psfig{file=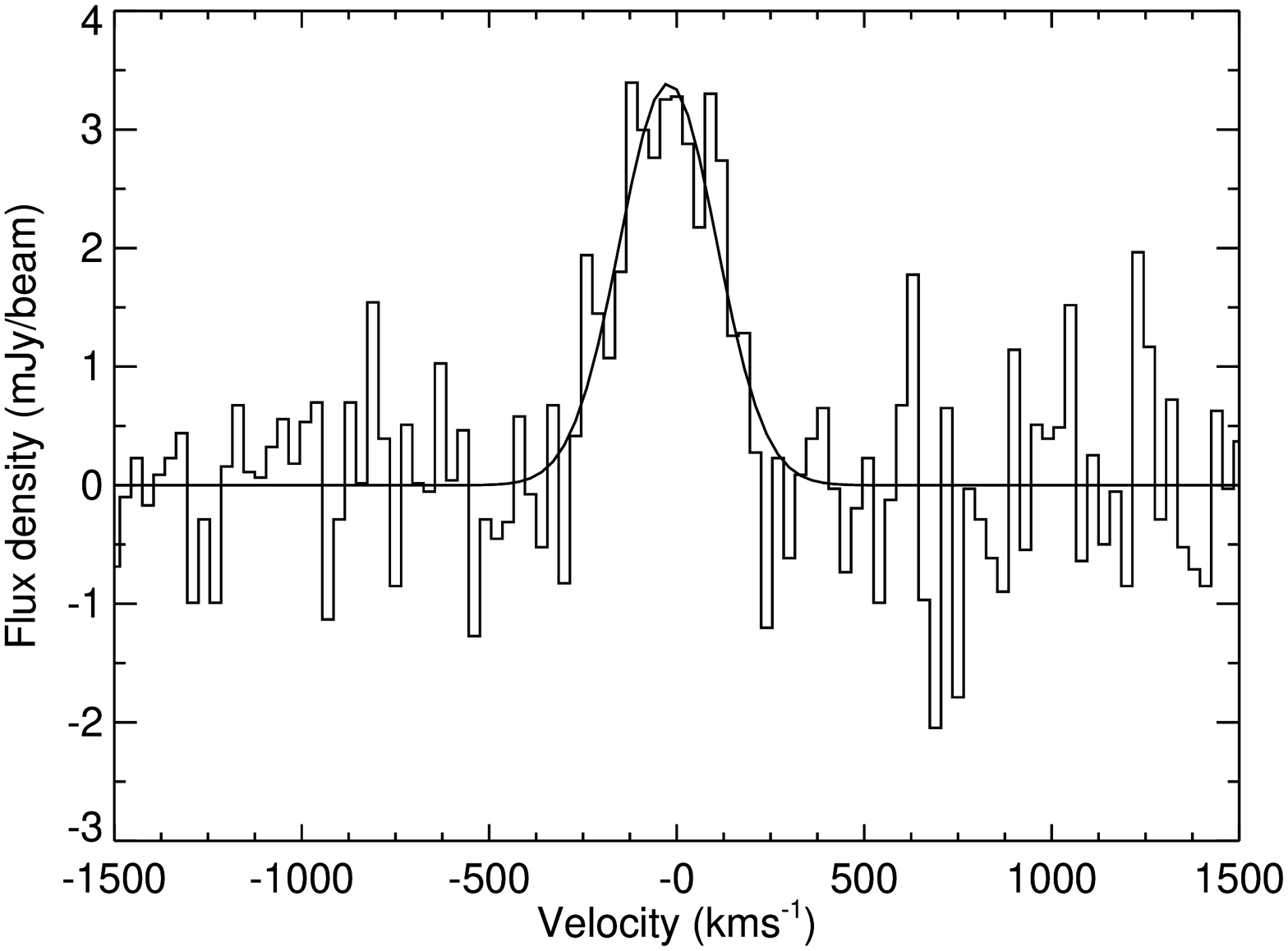, width=6.0cm, angle=90}
\label{fig:3mm}
}
\subfigure[CO(5-4)]{
\psfig{file=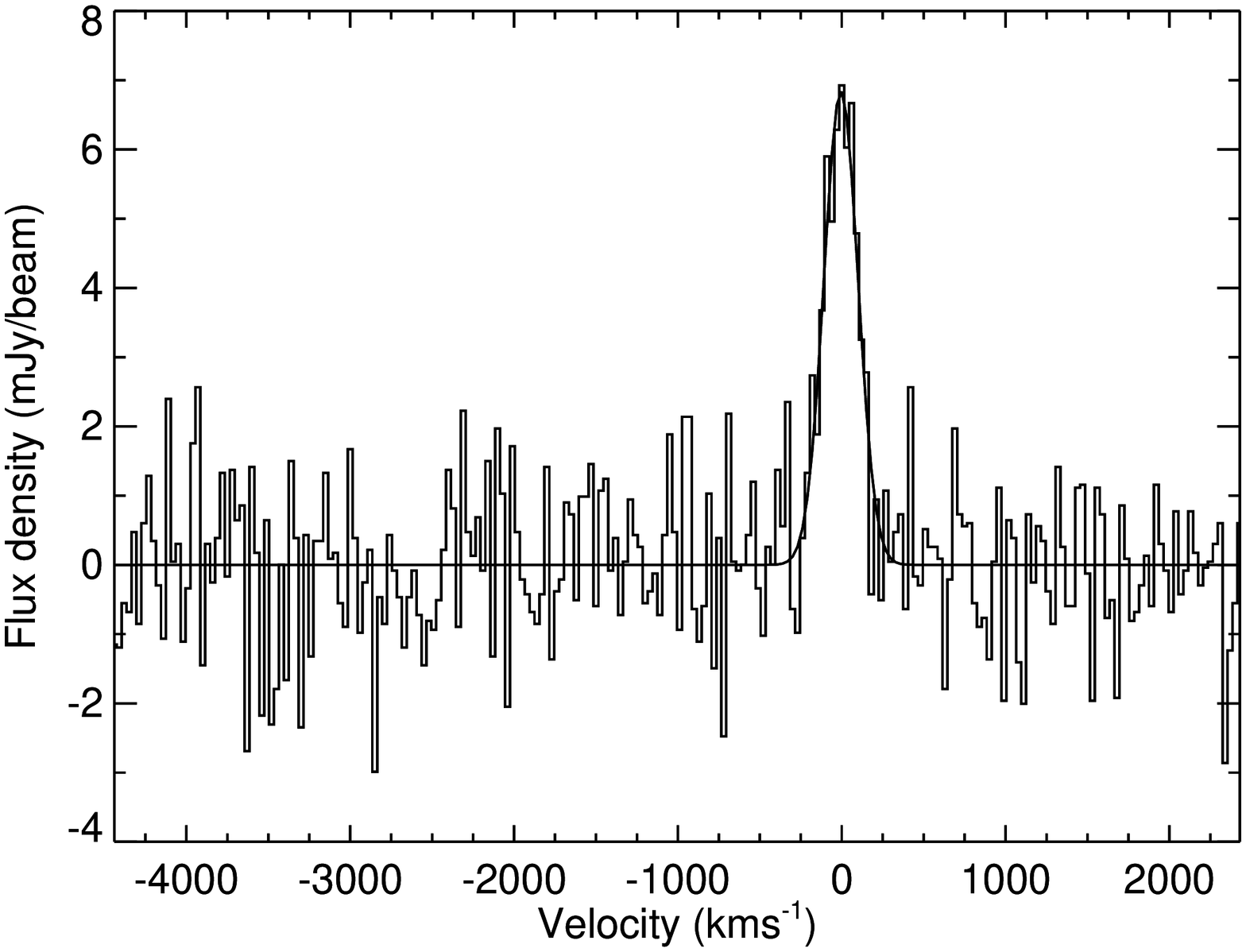, width=6.0cm, angle=90}
\label{fig:2mm}
}
\subfigure[\highCI\ and CO(7-6)]{
\psfig{file = 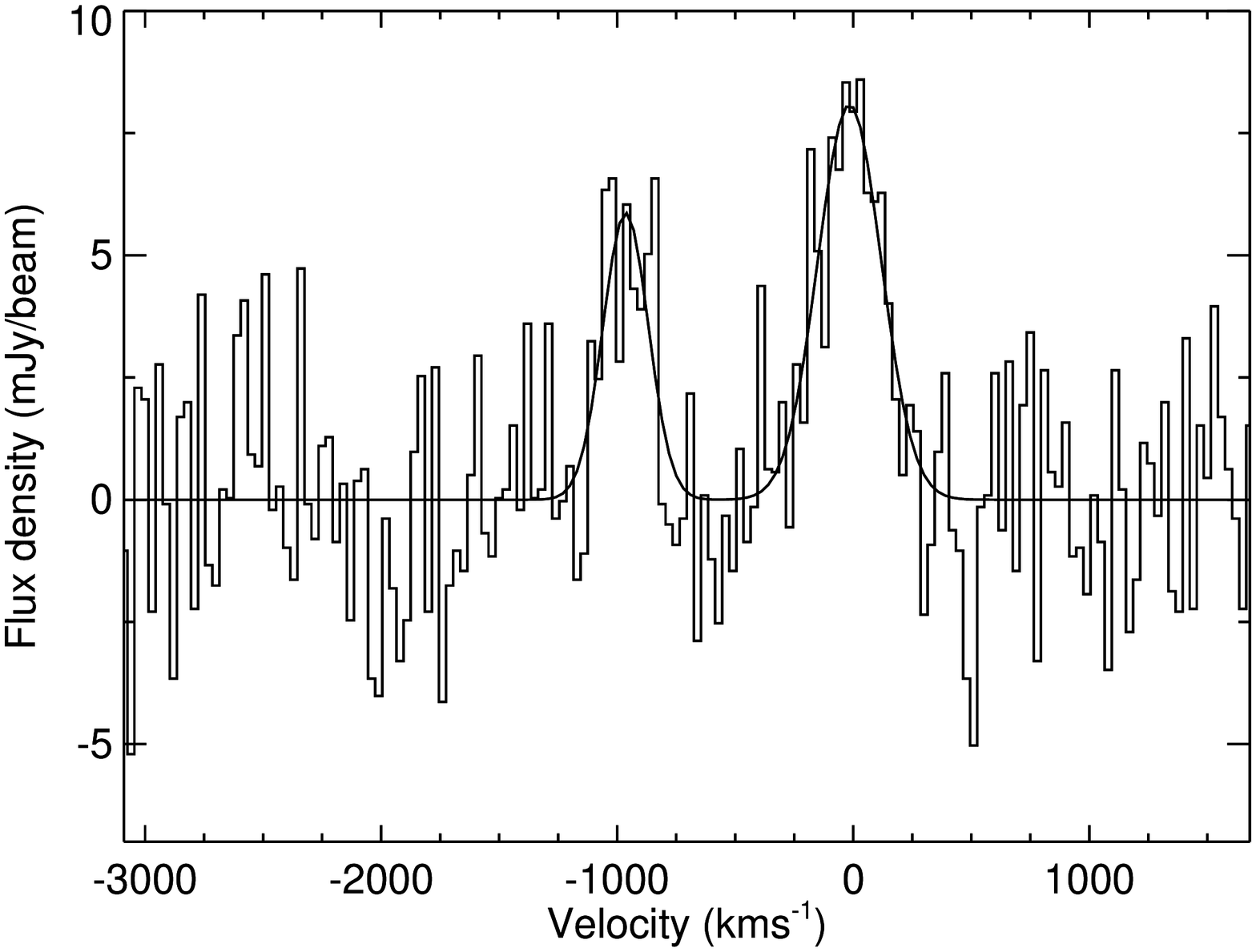, width=6.0cm, angle=90}
\label{fig:1mm}
}
\subfigure[\lowCI]{
\psfig{file = 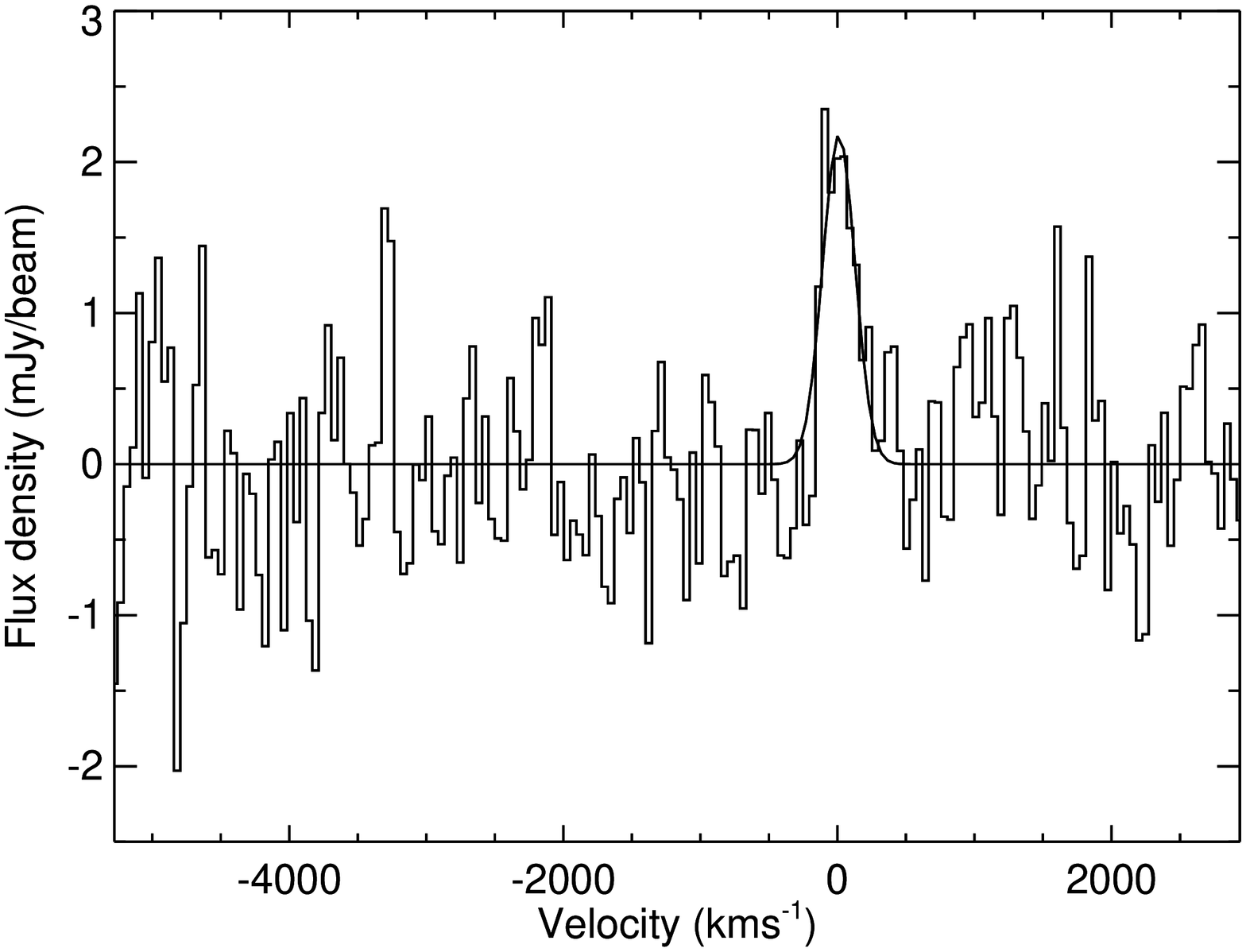, width=6.0cm, angle=90}
\label{fig:ci10}
}
\caption{\noindent Continuum subtracted CO molecular and [CI] atomic lines from the PdBI 3, 2 and 1.3 mm observations, the spectra are in 30 \kms\ bins unless otherwise stated. \textbf{(a):} The $3~\rm{mm}$ band PdBI spectrum of AMS12 with the narrow band correlator ($1~\rm{GHz}$ bandwidth) showing the CO(3-2) emission line. \textbf{(b):} Spectrum from the 2 mm band shows the CO(5-4) line. \textbf{(c):} Spectrum from 1.3 mm band shows \highCI\ and CO(7-6) (line at 0 \kms). \textbf{(d):} The \lowCI\ line detection, spectrum in 20 MHz bins. See Table \ref{tab:CO} for the line parameters derived from the Gaussian fits to these lines.}
\label{fig:123mm}
\end{figure*}

The data were reduced and analysed using the \textsc{gildas} software\footnote{http://www.iram.fr/IRAMFR/GILDAS} \citep{Guilloteau2000}. The final data cube for the 3 mm band observations achieved an rms noise of $0.70~\rm{mJy~beam^{-1}}$ per $30~\rm{kms^{-1}}$ channel. The resulting spectrum can be seen in Figure \ref{fig:3mm} \citep[as published in][]{Martinez2009}. For the 2 mm band an rms noise of $0.95~\rm{mJy~beam^{-1}}$ per $30~\rm{kms^{-1}}$ bin for the final data cube was reached (see Figure \ref{fig:2mm}). The 1.3 mm band's final data cube has an rms noise of $2.0~\rm{mJy~beam^{-1}}$ per $30~\rm{kms^{-1}}$ bin (Figure \ref{fig:1mm}). 

The 1.3 mm band observations revealed the \highCI\ line. This prompted an investigation into the detection of the \lowCI\ line in AMS12. These observations were carried out with the PdBI using 5 antennas in the D configuration over 7 days during July and August 2011. The observations were centred on 130.660 GHz ($\nu_{\rm{rest}} = 492.161$ GHz) which falls in the 2 mm band. A total of 12.3 hours of on-source integration time was obtained with varying weather conditions (PWV ranging from 5-11 mm). Data were flagged on the 19 July, 02 and 05 August due to receiver system temperatures exceeding $\sim 400$ K. The uv table was created from the \textit{WideX} correlator data with an rms noise of $0.577~\rm{mJy~beam^{-1}}$ per 20 MHz bin (see Figure \ref{fig:ci10}).   

The visibilities were imaged using natural weighting to maximise the point source sensitivity. We used the software's recommendations for the image and map sizes for the data-cube set up. The dirty images were cleaned using the Hogbom algorithm down to $\sim$ 3 times the thermal noise. The line spectra were extracted from a single pixel centred on the source's position. With the resolution offered by the compact configuration, the source is unresolved and so extracting the spectra from single pixels retains most, if not all, of the information from the source. 

\subsection{Results}
\begin{figure}
\begin{center}
\psfig{file=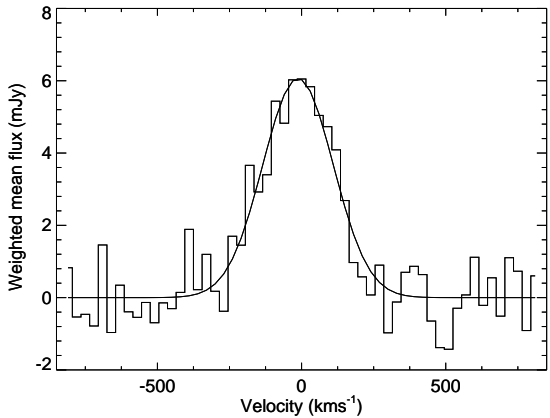, width=6.0cm, angle=90}
\caption{\noindent The average CO profile from the CO(3-2), CO(5-4) and CO(7-6) transitions.  The transitions are weighted by the inverse square of the noises.  The profile is fit with a Gaussian centred on -15 \kms and with a FWHM of $290~\pm~20$ \kms.}
\label{fig:stack}
\end{center}
\end{figure}

The linewidths for the three CO and two [CI] lines are $\sim 200-300$ \kms. The CO(3-2), CO(5-4) and CO(7-6) line emission towards AMS12 are detected at $11\sigma$, $17\sigma$ and $11\sigma$ significance respectively. Figure \ref{fig:stack} shows the stacked CO transition profiles with their baselines subtracted and weighted by the inverse squared noise.  The averaged CO lines are well fit with a Gaussian of FWHM of $290~\pm~20$ \kms\ and offset from $z=2.7668$ by $-15 \pm 9$ \kms\ agreeing with $z_{\rm{CO}}$ of $2.7672\pm0.0003$ as determined by the individual lines. 

The central frequency, flux density and full-width half maximum of each line were obtained by fitting a Gaussian to the data, using  least-squares minimisation which additionally allows an estimation of the uncertainty for the parameters. These are shown in Table \ref{tab:CO} along with other derived parameters. The line luminosities of both the CO transitions and the [CI] lines are given in both \lsol\ (L$_{\rm{CO}}$), and K \kms\ pc$^{2}$ (L$^{\rm{T}}_{\rm{CO}}$), \citep{Solomon1992}.  
%
%

An initial interpretation of the CO lines detected in AMS12, reveal they are subthermally excited, i.e. the line ratios are $<1$. The L$^{\rm{T}}_{\rm{CO(5-4)}}$/L$^{\rm{T}}_{\rm{CO(3-2)}}$ ratio is $0.59 \pm 0.05$, and the L$^{\rm{T}}_{\rm{CO(7-6)}}$/L$^{\rm{T}}_{\rm{CO(3-2)}}$ ratio $0.46 \pm 0.06 $.

\begin{table*}
\begin{minipage}{175mm}
\begin{tabular}{cccccccc}
      \hline
      \hline
Line & $\nu_{obs}$ & $\rm{S}_{\nu}$ & $\Delta\rm{V}_{FWHM}$ & $\rm{I_{CO}}$ & $\rm{V}^{\rm{a}}$ & $\rm{L}/10^{8}$ & $\rm{L^{T}}/10^{10}$  \\
& $\rm{[GHz]}$ & $\rm{[mJy]}$ & $\rm{[kms^{-1}]}$ & $\rm{[Jy~kms^{-1}]}$ & $\rm{[kms^{-1}]}$ & [\lsol] & $\rm{[K~kms^{-1}~pc^{2}]}$ \\
\hline
CO(3-2) & $91.803\pm0.004$ & $3.4\pm0.3$ & $300\pm30$ & $1.1\pm0.1$ & $-23\pm15$ &  $0.55\pm0.05$ & $4.2\pm0.4$ \\
CO(5-4) & $152.898\pm0.005$ & $6.9\pm0.4$ & $240\pm20$ & $1.8\pm0.1$ & $-5\pm10$ & $1.5\pm0.1$ & $2.5\pm0.2$ \\
CO(7-6) & $214.159\pm0.013$ & $8.1\pm0.7$ & $320\pm40$ & $2.7\pm0.3$ & $-16\pm20$ & $3.3\pm0.4$ & $1.9\pm0.2$\\
\lowCI\ & $130.657\pm0.009$ & $2.2\pm0.3$ & $280\pm50$ & $0.65\pm0.1$ & $11\pm20$ & $0.47\pm0.10.$ & $1.2\pm0.3$\\
\highCI\ & $214.837\pm0.016$ & $5.9\pm0.9$ & $230\pm30$ & $1.5\pm0.3$ & $36\pm20$ & $1.7\pm0.3$ & $1.0\pm0.2$\\
   \hline
   \hline
  \end{tabular}
  \caption{\noindent Observed $\rm{CO}$ and $\rm{CI}$ line parameters towards AMS12 derived from the individual Gaussian fits. $^{\rm{a}}$Velocities are reported relative to a redshift of 2.7668. } 
\label{tab:CO} 
\end{minipage}
\end{table*}

\subsection{Continuum measurements}

We use a compilation of data between 70~$\mu$m and 3.0 mm to infer the
best-fitting parameters for the FIR SED of AMS12. The fluxes are given in Table \ref{tab:FIRphot}.

\subsubsection{Existing data: MAMBO and Spitzer}

The Max-Planck Millimetre Bolometer Array (MAMBO) observations were done at a wavelength of 1.2 mm. The source was observed with other sources in the AMS sample in blocks of typically 20 minutes. The data were reduced using the \textsc{mopsic} pipeline. The rms noise achieved was $\sim 0.55$ mJy beam$^{-1}$ \citep[see][for details]{Martinez2009}.  

We use archival Spitzer measurements of AMS12 at 160 and 70 $\mu$m. These were made as part of the Spitzer extragalactic First Look Survey (FLS). See \citet{Frayer2006} for details on reduction and handling.

\subsubsection{New data: Herschel and PdBI}
The FLS field was observed as part of the Herschel Multi-tiered Extragalactic Survey (HerMES)\footnote{http://hermes.sussex.ac.uk} \citep{Oliver2010}.
Fluxes of the AMS objects at 250, 350 and 500 $\mu$m were measured off
the level 2 SPIRE mosaics distributed by the Herschel Science Archive.
Aperture photometry was carried out in apertures of 13, 17 and 23\arcsec\ at
250, 350 and 500 $\mu$m, respectively, with background annuli 23-60\arcsec,
30-100\arcsec\ and 40-140\arcsec. Aperture corrections were derived
from the beams associated with version 1.0 of the Spectral and Photometric Imaging Receiver (SPIRE) beam release
note \citep{Sibthorpe2010}, sampled with 1\arcsec\ pixels, and the
corresponding beam areas applied. The beam areas assumed were 426, 771
and 1626 square arcseconds at 250, 350 and 500 $\mu$m, respectively, and
aperture corrections were 1.51, 1.52 and 1.60, respectively.

The continuum towards AMS12 in the PdBI 3 mm band was found over the line free bandwidth $684.4~\rm{MHz}$ or $2053.8~\rm{kms^{-1}}$, leading to an rms noise of $0.09~\rm{mJy}$. For the $2~\rm{mm}$ band, $3.35~\rm{GHz}$ of the line free spectrum was fit yielding an rms noise of 0.06 mJy.  The 1.3 mm band yielded the continuum measurement over a spectral line free region spanning 2.572 GHz, giving an rms noise of 0.19 mJy.

\begin{table}
\begin{center}
 \begin{tabular}{|c|c|}
 \hline
 \hline
$\lambda_{\rm{obs}}$ ($\mu$m) & Flux (mJy)\\ \hline
75& $15.0\pm5$  \\
158& $74.0\pm20$ \\
250& $65.0\pm6$ \\
350& $49.4\pm5$\\
500&$42.5\pm12$\\
1200&$3.7\pm0.6$\\
1400&$2.16\pm0.19$\\
1960&$1.01\pm0.06$\\
3260&$0.75\pm0.09$\\
 \hline
\hline
 \end{tabular}
 \caption{The flux measurements of the FIR infrared measurements from observations by Spitzer/MIPS, Herschel/Spire, MAMBO and PdBI.}
 \label{tab:FIRphot}
\end{center} 
\end{table}  
 
\section{Modelling the FIR emission}

At FIR wavelengths dust is not optically thick, and the
SED of the radiation can be described by a
modified black body.  If the absorption coefficient of the dust,
$\kappa(\nu)$, is assumed to follow a law $\propto \nu^{\beta}$, the
emission will be given by:

\begin{equation}
L_{\nu} = \frac{A \nu^{3+\beta}}{(e^{\frac{h\nu}{k T_{\rm D}}}-1)}
\end{equation}

\noindent where $A$ is a normalisation term given by:

\begin{equation}
A = { L_{\rm FIR} \over \zeta(\beta + 4) \Gamma(\beta+4)  }{h \over k T_{\rm D}}.
\end{equation}

\noindent The three variables are: the dust temperature $T_{\rm D}$,
the emissivity index $\beta$ and the FIR-luminosity $L_{\rm
  FIR}$. Here $h$ and $k$ are Planck's and Boltzmann's constants,
respectively, while $\zeta$ and $\Gamma$ are the Riemann zeta function and the Gamma function, respectively. 

In order to determine the parameters of the fit to the FIR SED, we use Bayes' theorem:

\begin{equation}
 P(\{x_1...x_n\}|data) = \frac{P(data|\{x_1...x_n\})P(\{x_1...x_n\})}{P(data)}
\end{equation}

Here the $\{x_1...x_n\}$ incorporates the $n$ variable parameters of the model used. In this case, the graybody model with parameters T$_D$, $\beta$ and \lfir\ fitted. The $data$ are the observed fluxes. 

The posterior probabilities $P(\{x_1...x_n\}|data)$ may be split into each parameter of interest through marginalisation. This is done by integrating the posterior probability density function (PDF) over the other parameters. Rewriting Bayes' theorem with the parameters under consideration explicitly stated and $S_{\nu}$ as the observable data we get,

\begin{equation}
 P(T_D,\beta,L_{\rm{FIR}}|S_{\nu}) \propto P(S_{\nu}|T_{D},\beta, L_{FIR})P(T_{D})P(\beta)P(L_{\rm{FIR}})
\label{eqn:bayes}
\end{equation}
    
\noindent where the evidence is treated as a normalisation term. The likelihood $P(S_{\nu}|T_{D},\beta, L_{\rm{FIR}})$, is given by a Gaussian distribution:
\begin{equation}
P(S_{\nu}|T_{D},\beta, L_{\rm{FIR}}) \propto e^{-\sum_i\left(S_{\nu i} - S_{\nu,m}(T_{D},\beta,L_{\rm{FIR}}) \over \sqrt{2}\sigma_{i} \right)^{2}}
\end{equation} 
\noindent with $S_{\nu,m}(T_{D},\beta, L_{\rm{FIR}})$ being the predicted flux given T$_{\rm{D}}$, $\beta$ and $L_{\rm{FIR}}$. Assuming Gaussian errors, maximising the likelihood is equivalent to minimising the $\chi^2$ statistic where $\chi^2  = \sum_i\left(S_{\nu i} - S_{\nu,m}(T_{D},\beta,L_{\rm{FIR}})\over \sigma_{i}\right)^2$.

In order to get the posterior PDFs for a particular parameter we can marginalise $P(T_{D},\beta,L_{\rm{FIR}}|S_{\nu})$ over the other two parameters. For example, to get $P(T_{D}|S_{\nu})$ we integrate over $\beta$ and $L_{\rm{FIR}}$.

\begin{equation}
 P(T_{D}|S_{\nu}) = \iint P(T_{D},\beta, L_{\rm{FIR}}|S_{\nu}) d\beta~dL_{\rm{FIR}}
\end{equation}

\noindent and similarly for $P(\beta|S_{\nu})$ and $P(L_{\rm{FIR}}|S_{\nu})$.

\begin{figure*}
\begin{center}
\subfigure[]{
\psfig{file=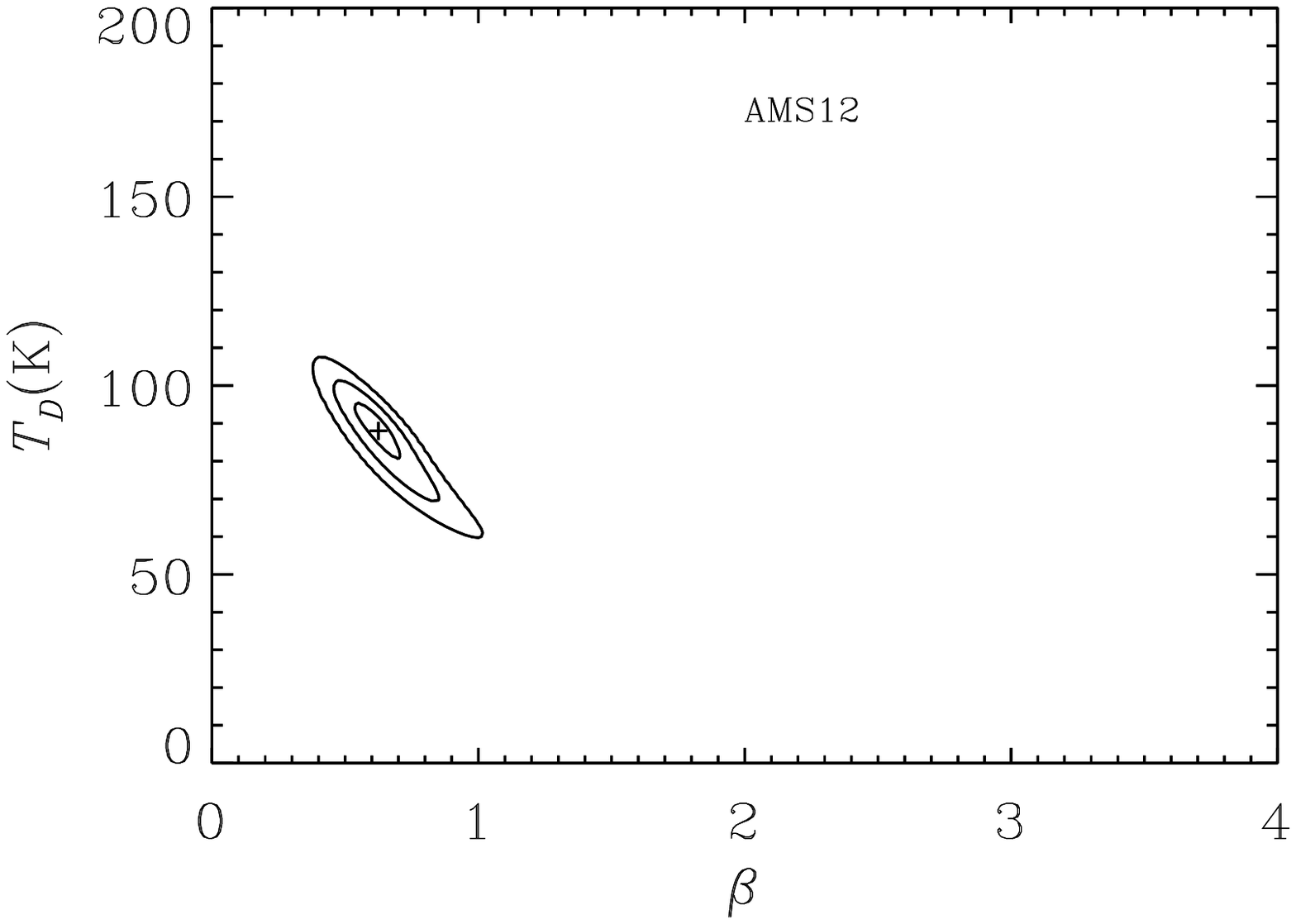, width=8cm, angle=0}
\label{fig:dust_conts}
}
\subfigure[]{
\psfig{file=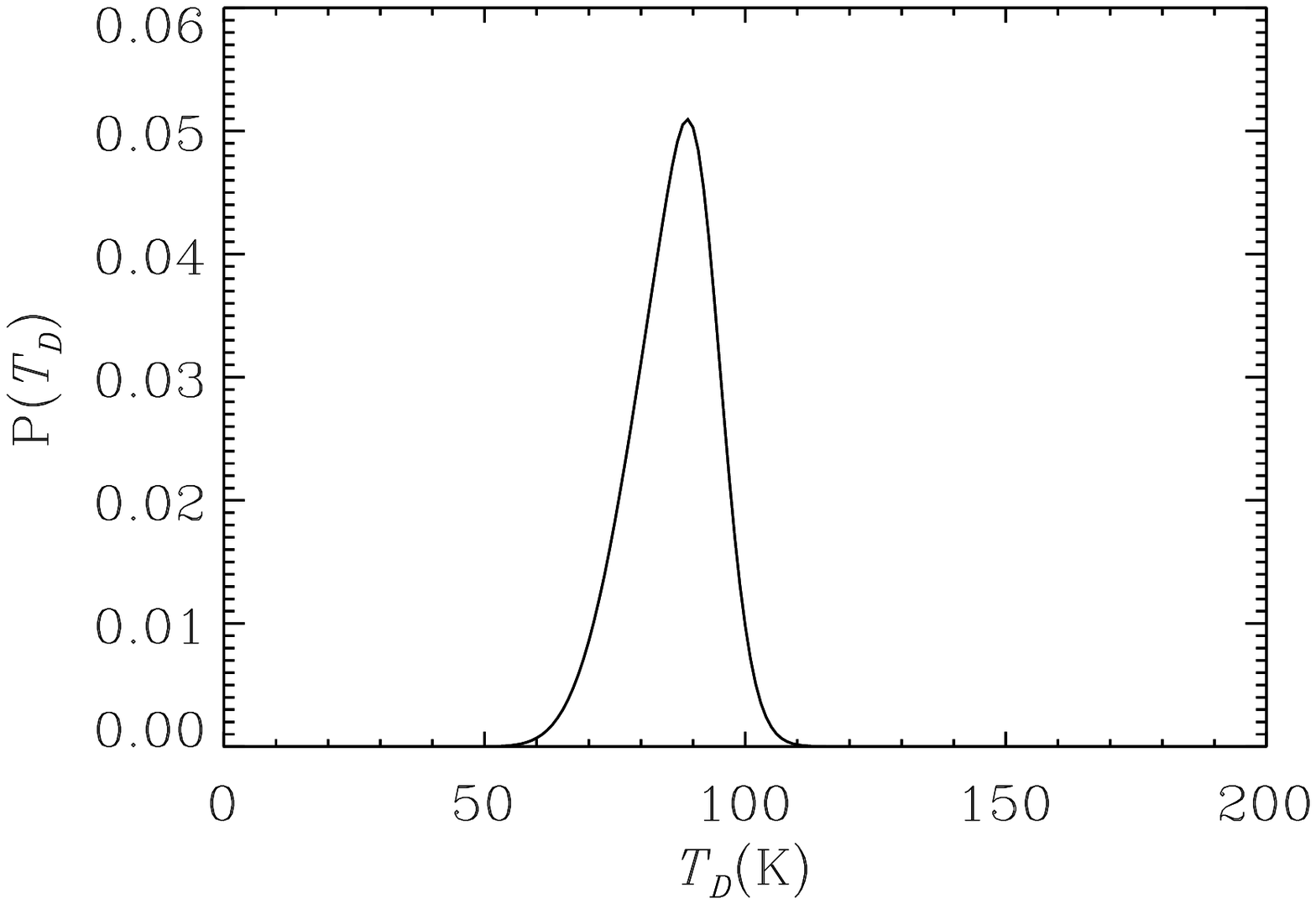, width=8cm, angle=0}
\label{fig:dust_temp}
}
\subfigure[]{
 \psfig{file = 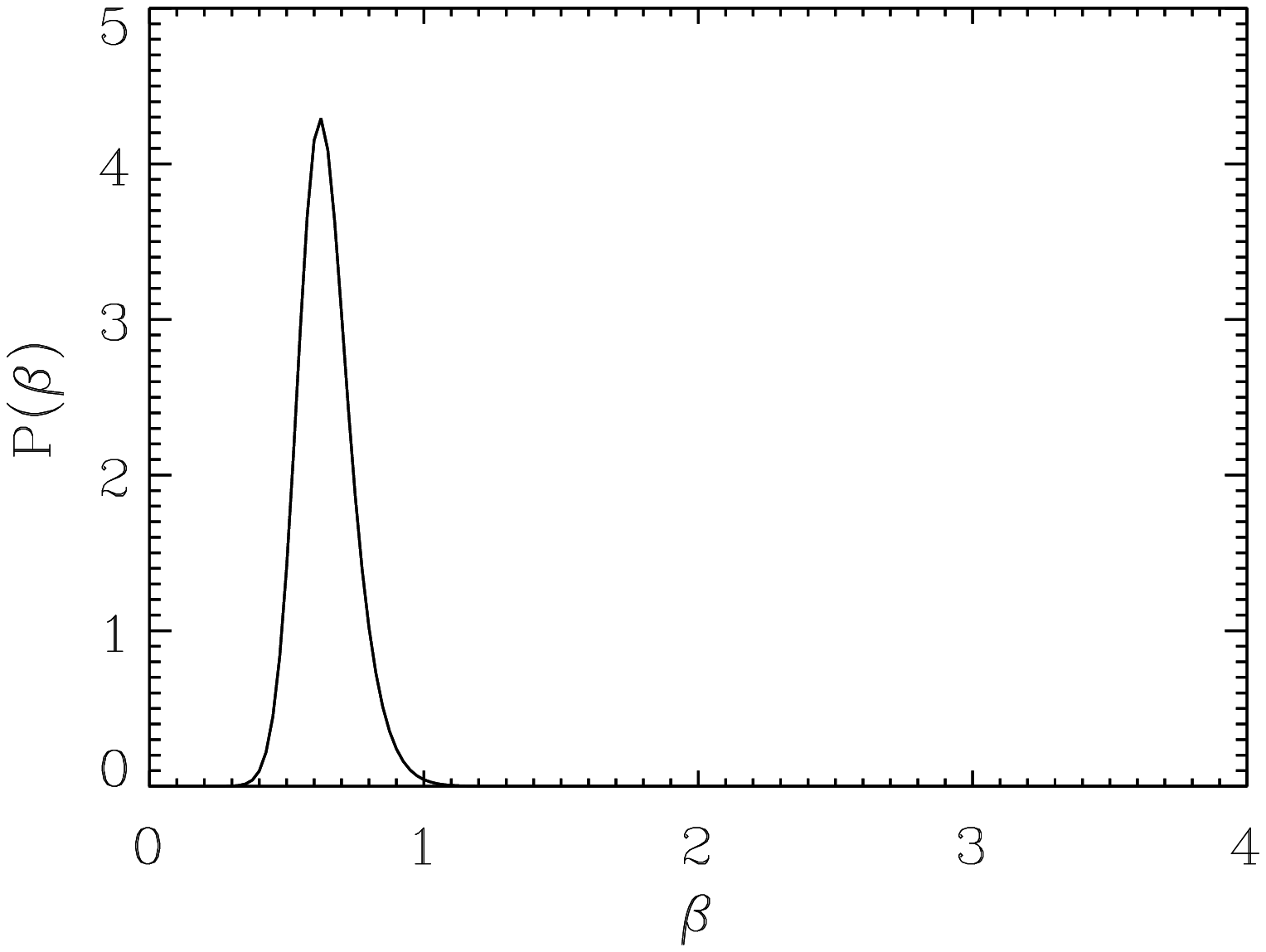, width=8cm, angle=0}
 \label{fig:dust_beta}
 }
 \subfigure[]{
 \psfig{file=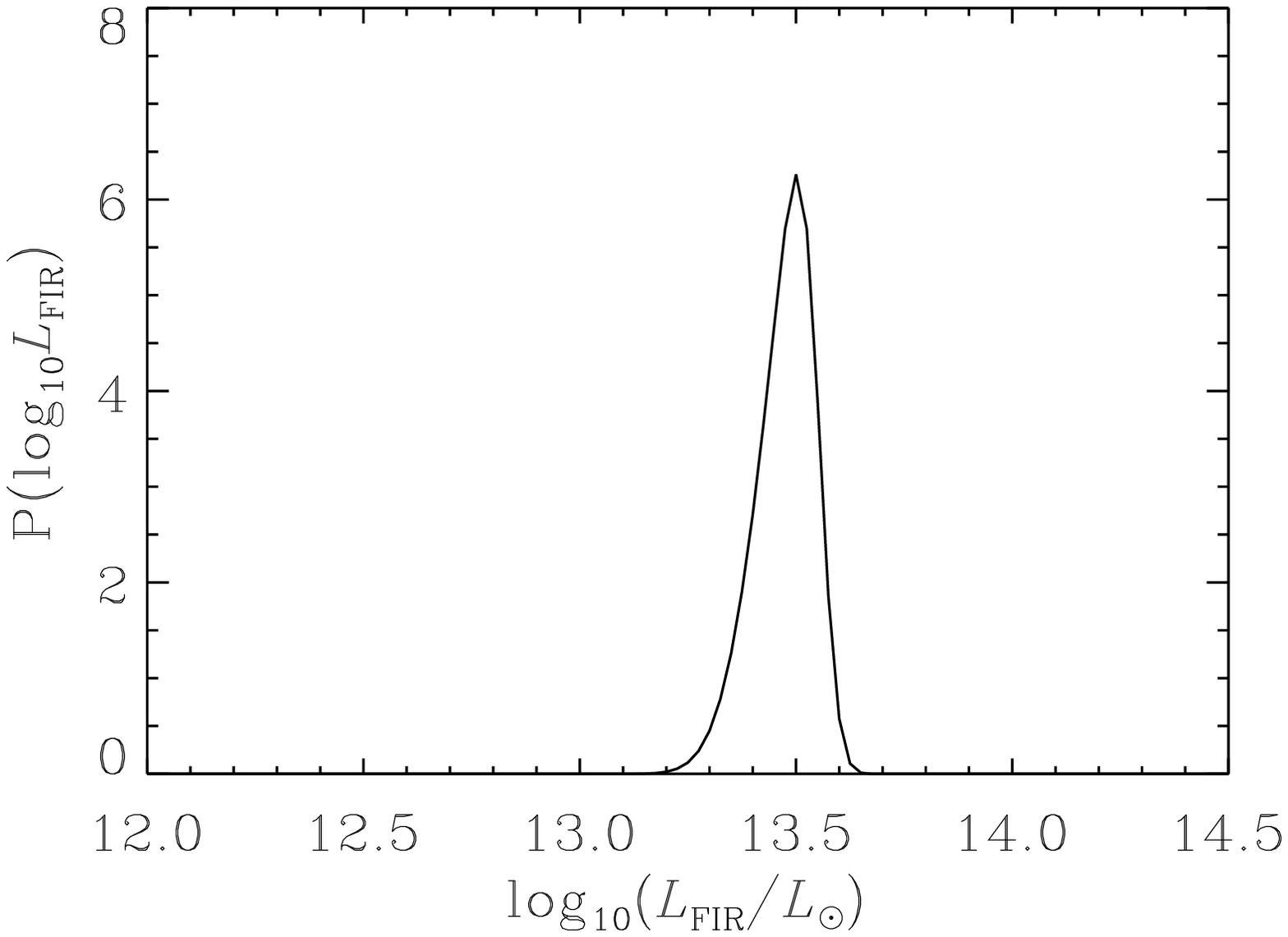, width=8cm, angle=0}
 \label{fig:dust_lfir}
 }
\caption{\noindent \textbf{(a):} The contours of the dust temperature versus the emissivity index for AMS12 marginalised over \lfir. The best fits to the temperature and emissivity index correspond to 88 K and 0.6. \textbf{(b):} The posterior PDF for the dust temperature, marginalised over $\beta$ and \lfir, with best fit $89\pm8$ K. \textbf{(c):} The posterior PDF for the $\beta$ parameter, marginalised over T$_{\rm{D}}$ and \lfir, with best fit value $0.6\pm0.1$. \textbf{(d):} The posterior PDF of the L$_{FIR}$, marginalised over T$_{\rm{D}}$ and $\beta$, the best fit is $\rm{log_{10}}(\rm{L}_{FIR}/$\lsol)$ = 13.5 \pm0.1$. }
\label{fig:dust_PDFs}
\end{center}
\end{figure*}

Figure \ref{fig:dust_conts} shows the contours of dust temperature and emissivity index, marginalised over the FIR luminosity, $P(T_{D},\beta|S_{\nu})$.  Figures \ref{fig:dust_temp}, \ref{fig:dust_beta} and \ref{fig:dust_lfir} show the posterior distribution function for the dust temperature, emissivity index and \lfir\ marginalised over the other two parameters respectively. 

\subsection{Far-infrared luminosity}

\begin{figure} 
\begin{center}
\psfig{file=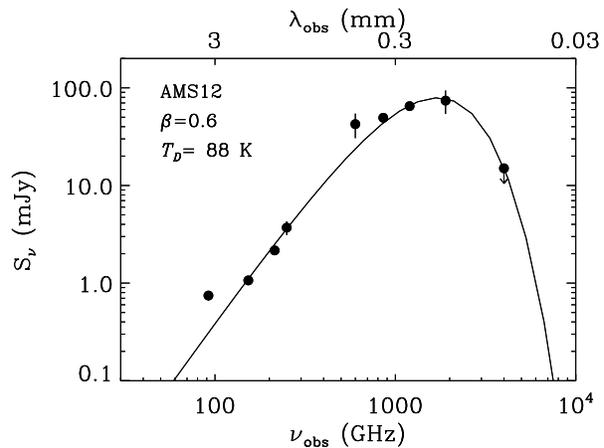, width=8cm, angle=0}
\caption{\noindent The graybody dust model fit to AMS12's FIR SED. The points are from observations from PdBI, MAMBO, Herschel/Spire and Spitzer/MIPS. The best fitting temperature and spectral index are shown on the figure.}
\label{fig:dust}
\end{center}
\end{figure}

Figure \ref{fig:dust} shows the FIR SED for AMS12. The best fitting temperature is T$_{D}~=~88 \pm 8$ K and emissivity index of $\beta~=~0.6 \pm 0.1$. There are two points which are not well fit by the model; the 3 mm PdBI measurement, and the 500 $\mu$m Herschel point. The 3 mm point could possibly suffer from contamination from the radio continuum from the AGN, AMS12 has a steep extended radio spectrum and a flatter, inverted spectrum within a compact 150 pc region. Extrapolating the radio continuum out to $\sim90$ GHz using the flatter compact spectral index ($\alpha = -0.22$ where S$_{radio}\propto \nu^{-\alpha}$), \citep{Klockner2009}, shows a significant amount of the measured flux could be due to radio emission from the AGN. 

The 500 $\mu$m measurement could be boosted by FIR emission lines. \citet{Smail2011} have estimated the FIR emission lines may contribute $\grtsim20-40\%$ to the broad-band flux, thus there could be possible contamination from the [CII] ($\nu_{rest} = 158~\mu$m) line which at $z=2.7672$ is at 595 $\mu$m, \citep[the Herschel/Spire 500 $\mu$m band has a width of $\lambda/\Delta\lambda = 3$,][]{Griffin2006}.
 
Figure \ref{fig:dust} illustrates the importance of obtaining points at both the longer and shorter wavelengths: it is necessary to probe wavelengths shorter than those where the peak of the emission appears, to constrain the location of this peak and hence the temperature. In order to constrain the emissivity index, long wavelength data are critical, in this case the data at 1.2 mm from MAMBO, and the additional longer wavelength data from PdBI.

The best fitting \lfir\ determined from the single graybody model fitting of the FIR SED is $\rm{log_{10}}(\rm{L}_{FIR}/$\lsol)$ = 13.5\pm0.1$. The \lfir\ can be used to determine the star formation rates (SFRs) of the galaxies. Assuming that the \lfir\ is solely due to star formation, with young OB stars the main source of heating, we can determine the SFR from the \citet{Kennicutt1998} conversion,

\begin{equation}
 SFR = 1.7 \times 10^{-10}L_{IR} 
\end{equation}

In order to obtain the total SFR, the assumption of an initial mass function (IMF) is required. Here, the \citet{Salpeter1955} IMF is assumed. 

The SFR for AMS12, given the \lfir\ determined from the best dust fitting model, is $\sim 5300$ \msolyr. SFRs of this scale are seen in only the most extreme starburst galaxies \citep[see][for a review]{Chapman2005,Tacconi2006, Coppin2008,SolVan2005}. 

However, the typical temperature found for star forming galaxies in both the local Universe and at high redshift, is $\lesssim50$ K \citep[e.g][]{Farrah2003,Kovacs2006,Elbaz2011}. The temperature derived from the FIR SED of AMS12 is significantly higher than this.  It is possible that a significant fraction of the \lfir\ is due to heating from the AGN, which would therefore mean the SFR determined here is an overestimate. 

\subsection{Dust mass}
The mass of the FIR dust is also found from the \lfir\ by \citep[e.g.][]{Beelen2006};
\begin{equation}
M_{D} = \frac{L_{\rm{FIR}}}{4\pi\int\kappa_{(\nu_{rest})}B_{\nu_{rest}}(T_{D})d\nu}
\end{equation}
\noindent where $\nu_{rest}$ is the rest frame frequency found by $(1+z)\nu_{obs}$. The mass absorption coefficient $\kappa_{(\nu_{rest})} = \kappa_{(\nu_{obs})}(\nu_{rest}/\nu_{obs})^{\beta}$ is given by a power law. The mass absorption coefficient is the main source of uncertainty for the dust mass. We assume two different reference values of $\kappa_{(\nu_{obs})}$; first, $\kappa_{(\rm{250GHz})}=0.04$ m$^{2}$kg$^{-1}$ \citep[i.e.][]{Alton2004}, and secondly we use $\kappa_{(2400GHz)} = 2.64$ m$^{2}$kg${-1}$ \citep{Dunne2003}. $B_{\nu}(T)$ is the Planck function for a given temperature and frequency. 

\begin{equation}
B_{\nu}(T) = \frac{2h\nu^{3}}{c^{2}}\frac{1}{[e^{(\frac{h\nu}{kT})}-1]}
\end{equation}  

The dust mass for AMS12 using the fiducial values $T_{D} = 88 $ K and $\beta=0.6$ is $M_{D} \sim 1.6\times 10^{9}$ \msol\ using $\kappa_{(250GHz)}=0.04$ m$^{2}$kg$^{-1}$, and  $M_{D} \sim 9.2\times 10^{7}$ \msol\ using $\kappa_{(2400GHz)} = 2.64$ m$^{2}$kg$^{-1}$. The difference between these two estimates highlights the uncertainty in obtaining reliable dust masses. We can only estimate the dust mass to an order of magnitude. This is mostly due to the uncertainty in $\kappa$ and the different frequencies at which it is derived (note we have simply chosen two illustrative values here), however other uncertainties are also large. For example the $\sim 10\%$ uncertainty in $T_{D}$ alone contributes a relative uncertainty of $\sim30\%$ in the dust mass.

\section{Carbon Monoxide}

The detection of multiple rotational transition lines of CO towards AMS12 gives the CO ladder. The excitation of CO to higher rotational levels depends upon the ambient temperature and the density of the gas. These lines are optically thick which makes it difficult to infer the physical properties of the gas directly. 

\subsection{LVG modelling}

In order to model the line intensities of the rotational transitions of CO, coupled equations of radiative transfer and statistical equilibrium must be solved. A simplification which localises the problem, comes in the form of the LVG approximation. This assumes that there are large velocity gradients across the area of the gas which are significantly greater than local thermal velocities. These act to locally trap the photons emitted from the de-excitation of the CO molecule from the $J$ state to $(J-1)$. The probability that the photon will escape the region is given by the escape probability which depends upon the optical depth of the transition. The photon trapping and the escape probability of a transition act to raise the intensity of the line transition above the background intensity (i.e. from the cosmic microwave background radiation).  

We refer the reader to \citet{Scoville1974} and \citet{Goldreich1974} for a complete derivation of the model in a collapsing spherical molecular cloud. 

We have used a LVG code developed and kindly provided by C. Henkel. The LVG calculations require the collision coefficients of the molecules under consideration, in addition to a few input parameters.  These are, an ortho-to-para ratio for H$_{2}$, the temperature of the background CMB radiation, a chemical abundance of the molecule relative to H$_{2}$ and the velocity gradient.  The free parameters are the gas kinetic temperature, T$_{\rm{G}}$, and the overall density of molecular hydrogen, n(H$_{2}$). 

We have used single-component LVG models using the collision rates from \citet{Flower2001}, to investigate the CO excitation. In all calculations we used the H$_2$ ortho-to-para ratio of 3:1, and a cosmic microwave background temperature of T$_{\rm{CMB}}= 10.28~\rm{K}$ (corresponding to the redshift $z=2.767$). We adopted the fixed CO abundance per velocity gradient value of $[\rm{CO}]/(dv/dr) = 1 \times 10^{-5}~\rm{pc / kms^{-1}}$ \citep[e.g.][]{Weiss2005b,Weiss2007}. 

The LVG model provides the brightness temperatures T$_{\rm{b}}$, of each rotational transition from $J = 1$ to $J=11$, which can be compared to the observed flux densities by,

\begin{equation}
 S_{\rm{CO}} = \Omega \frac{\rm{T}_{\rm{b}}}{(1+z)}\frac{2k\nu_{obs}^{2}}{c^{2}}
\label{eqn:sco}
\end{equation}

\noindent where $\Omega$ is the source solid angle \citep[e.g.][]{Weiss2007}. Due to the fact that we have not resolved the source, $\Omega$ is kept as a free parameter. We use the equivalent source radius $r_0$ which is given by $r_0~=~D_{A}\sqrt{\Omega/\pi}$, where $D_{A}$ is the angular diameter distance.

With LVG models, there is a noted degeneracy between the parameters T$_{\rm{G}}$ and n(H$_{2})$ \citep[e.g.][]{Weiss2007,Ao2008}. The degeneracy arises from the dependency of the line optical depth on the level populations which in turn depend upon the values of T$_{\rm{G}}$ and n(H$_{2}$). 

In order to counteract this degeneracy and constrain the density and temperature further, we can use the information we have obtained from the continuum observations. We present the results from the LVG modelling below, firstly assuming no prior knowledge on the parameters, and then applying a prior on the temperature from the dust analysis.

\subsection{CO LVG model results}

\subsubsection{Assuming no prior knowledge of the gas kinetic temperature}

\begin{figure*}
\begin{center}
\subfigure[]{
\psfig{file=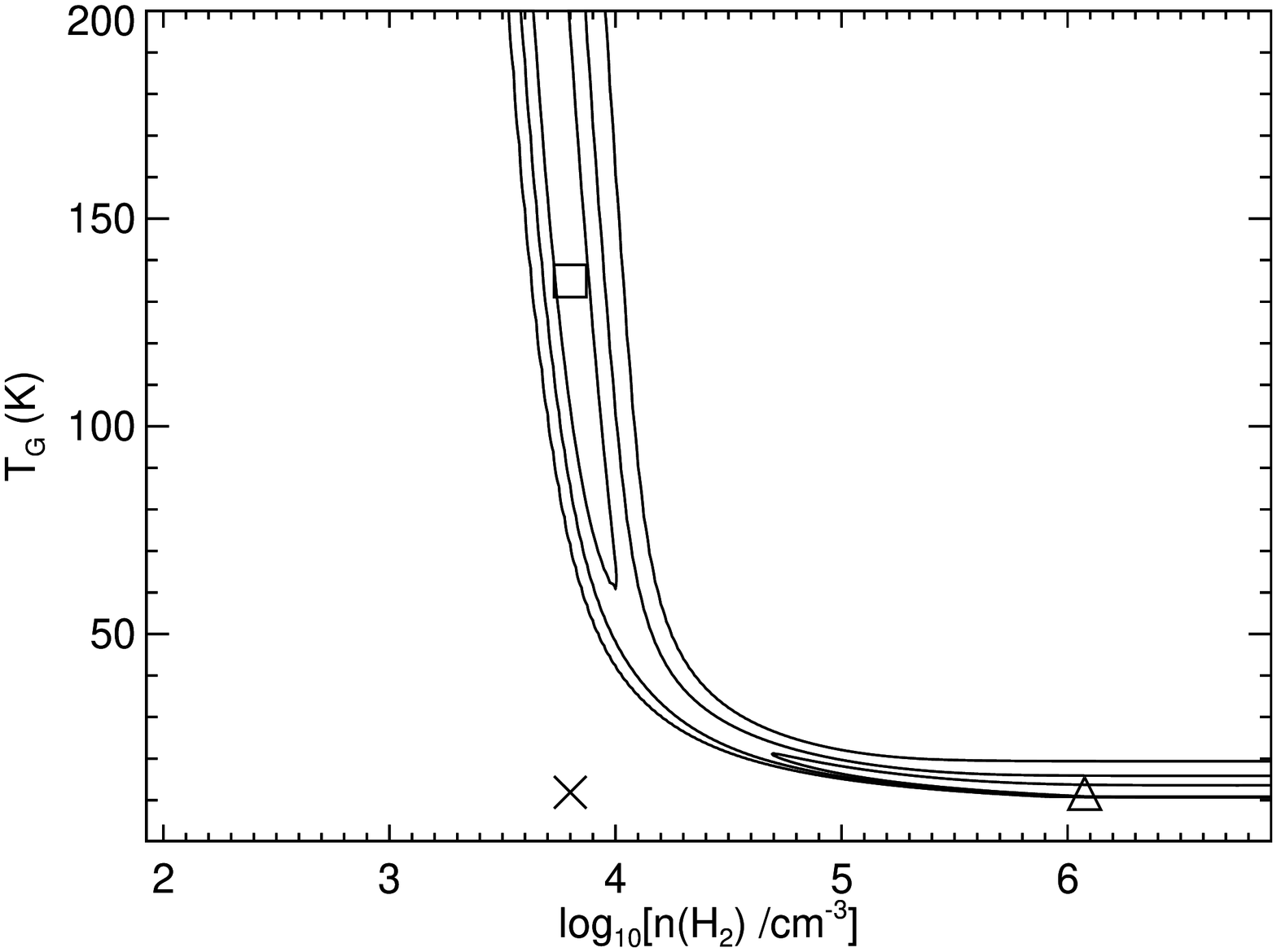, width=5.5cm, angle=90}
\label{fig:gas_cont}
}
\subfigure[]{
\psfig{file=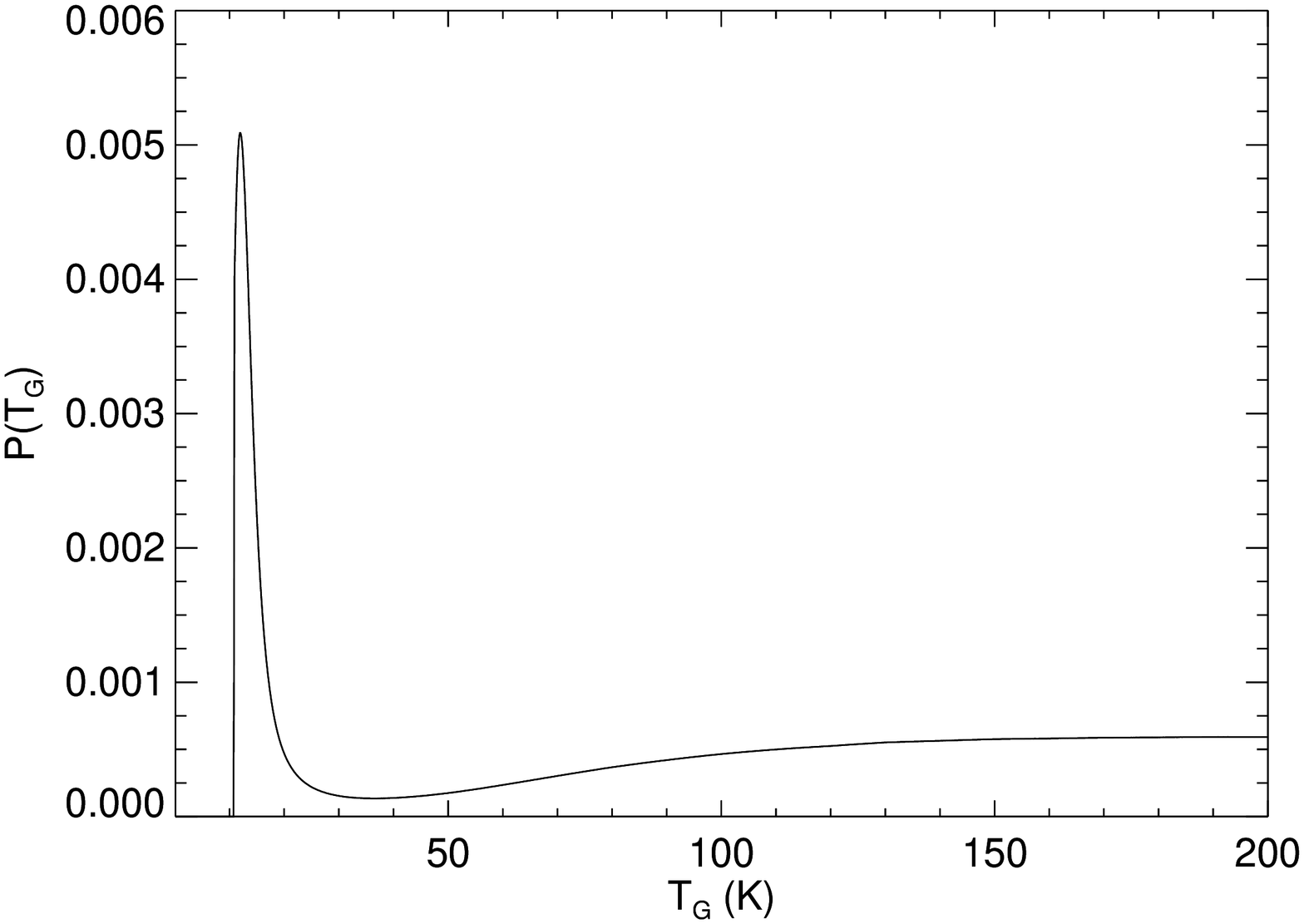, width=5.5cm, angle=90}
\label{fig:marg_T}
}
\subfigure[]{
 \psfig{file=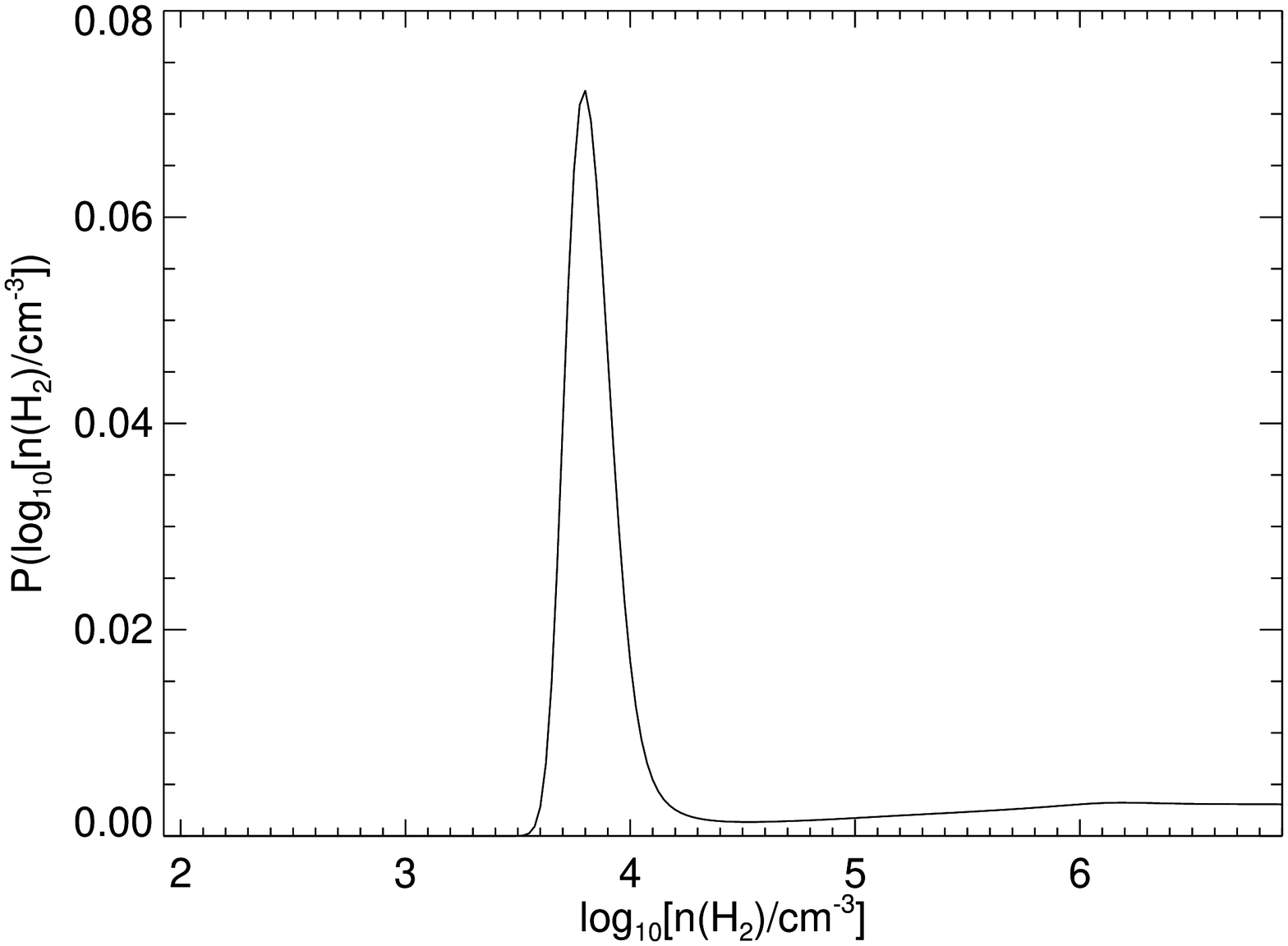, width=5.5cm, angle=90}
 \label{fig:marg_N}
}
\subfigure[]{
\psfig{file =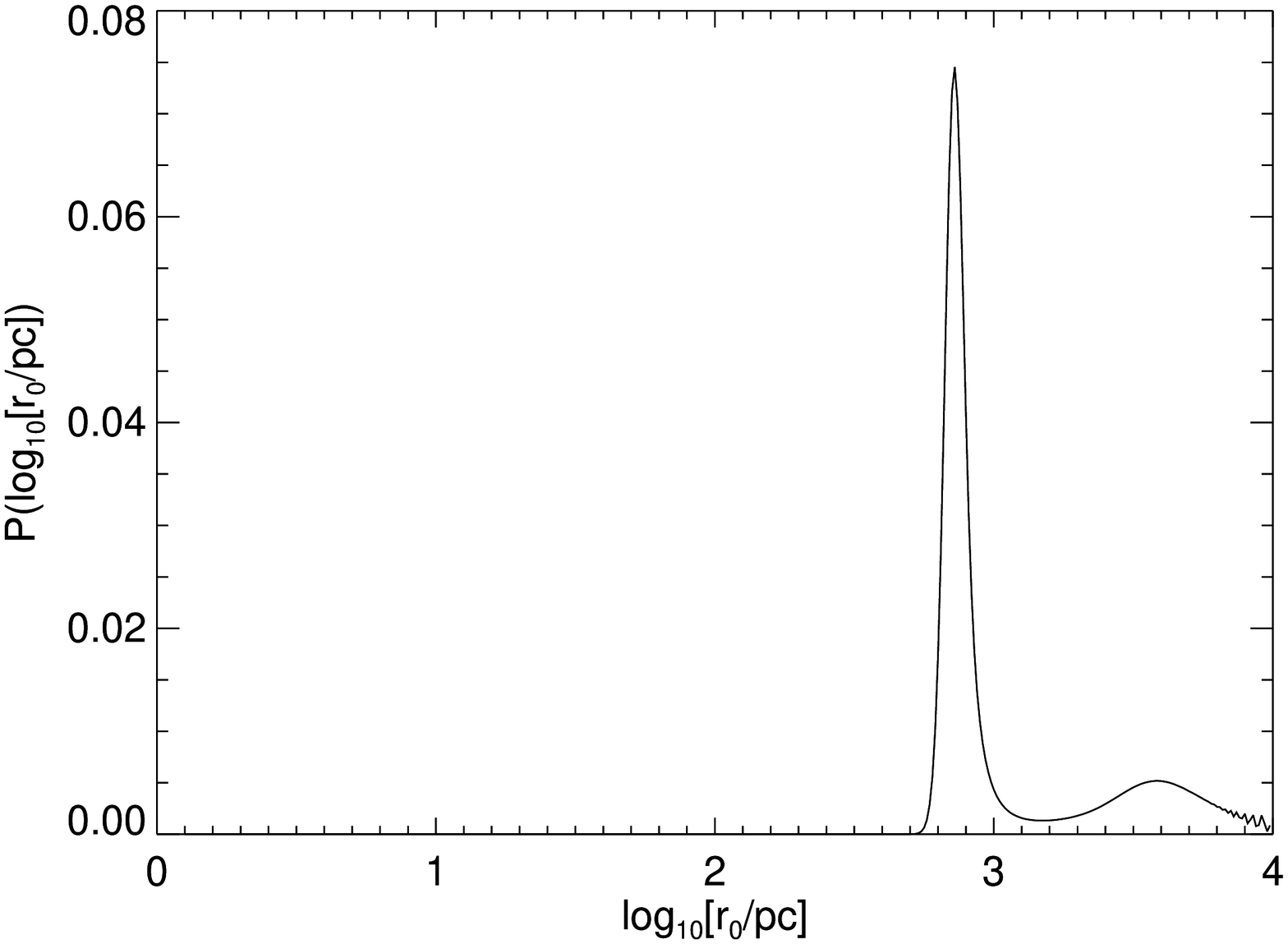, width=5.5cm, angle=90}
\label{fig:marg_r0}
}
\caption{\noindent \textbf{(a):} The contours of the temperature and density PDFs marginalised over the size. The open triangle marks the model of least $\chi^{2}$, while the open square marks an arbitrary spot in the higher temperature-lower density/size region, within the 1$\sigma$ confidence interval. The best values of the marginalised individual parameters are marked with an `X'. \textbf{(b):} The PDF of the gas temperature marginalised over the density and size parameters. \textbf{(c):} The PDF of the gas density parameter marginalised over the temperature and the size. \textbf{(d):}The PDF of the size of the emitting region of gas marginalised over the temperature and density.}
\label{fig:cont_plots}
\end{center}
\end{figure*}

The results from the LVG modelling using flat priors for all the parameters, show there is no single conclusive region of parameter space which points to the best fitting region as can be seen in Figure \ref{fig:cont_plots}. The marginalised PDFs of the individual parameters are also shown in Figure \ref{fig:cont_plots}. 

The minimum temperature ``floor'' in Figure \ref{fig:gas_cont} corresponds to the temperature of the background radiation at this redshift. The minimum density ``wall'' reflects the minimum density required to excite the higher CO lines to the observed levels. Figure \ref{fig:gas_cont} shows there are two regions which are within the 1$\sigma$ contour; the region surrounding the triangle and the region around the square.

Taking just our current CO measurements, we cannot confidently rule out either region. However, we have an upper limit of $\sim15$ kpc on the extent of the CO(7-6) emission from the unresolved PdBI 1.3 mm observations. The sizes which correspond to these temperatures and densities, are all physically possible, with the higher-temperature/lower-density solutions having lower sizes of the order of 1 kpc and less. The low-temperature/high-density solutions have radii of a few kpc. Figures \ref{fig:marg_T}, \ref{fig:marg_N}, and \ref{fig:marg_r0} show the marginalised probabilities of the unknown parameters. The peaks of these individual PDFs correspond to a T$_{\rm{G}}$ = \margflatT\  K, density of n(H$_2$) = \margflatN\ cm$^{-3}$ and $r_0$ = \margflatr\ kpc, the combination of these individually marginalised values are marked by an `X' in Figure \ref{fig:gas_cont}. They do not provide a good fit to the data.

The LVG model solution which corresponds to the lowest $\chi^{2}$ value has a T$_{\rm{G}}$ = \bestfitT\ K, a density of n(H$_2$) = \bestfitN\ cm$^{-3}$ and the size of the emitting region $r_0$ = \bestfitr\ kpc. This is marked with a triangle in Figure \ref{fig:gas_cont}. 

The low kinetic temperature of the gas from this solution is only slightly above the temperature of the background radiation at this redshift. In order to get the line intensities above the background radiation level, this solution has a high gas density. The optical depths of the lines from $J=1$ to $J=7$ are $\gg 1$. At high optical depths, the photons emitted in the de-excitation of the levels remain in the region longer i.e. do not escape. They are able to interact further, driving up the number of molecules in these excited states.

While the opacity of the lines is high, the collisional excitation and de-excitation processes in this model dominate over the radiative processes and the level populations are in actual thermal equilibrium. The excitation temperatures of these line transitions is equal to the kinetic gas temperature since the system is in local thermodynamic equilibrium (LTE).  

\begin{figure}
\begin{center}
\psfig{file=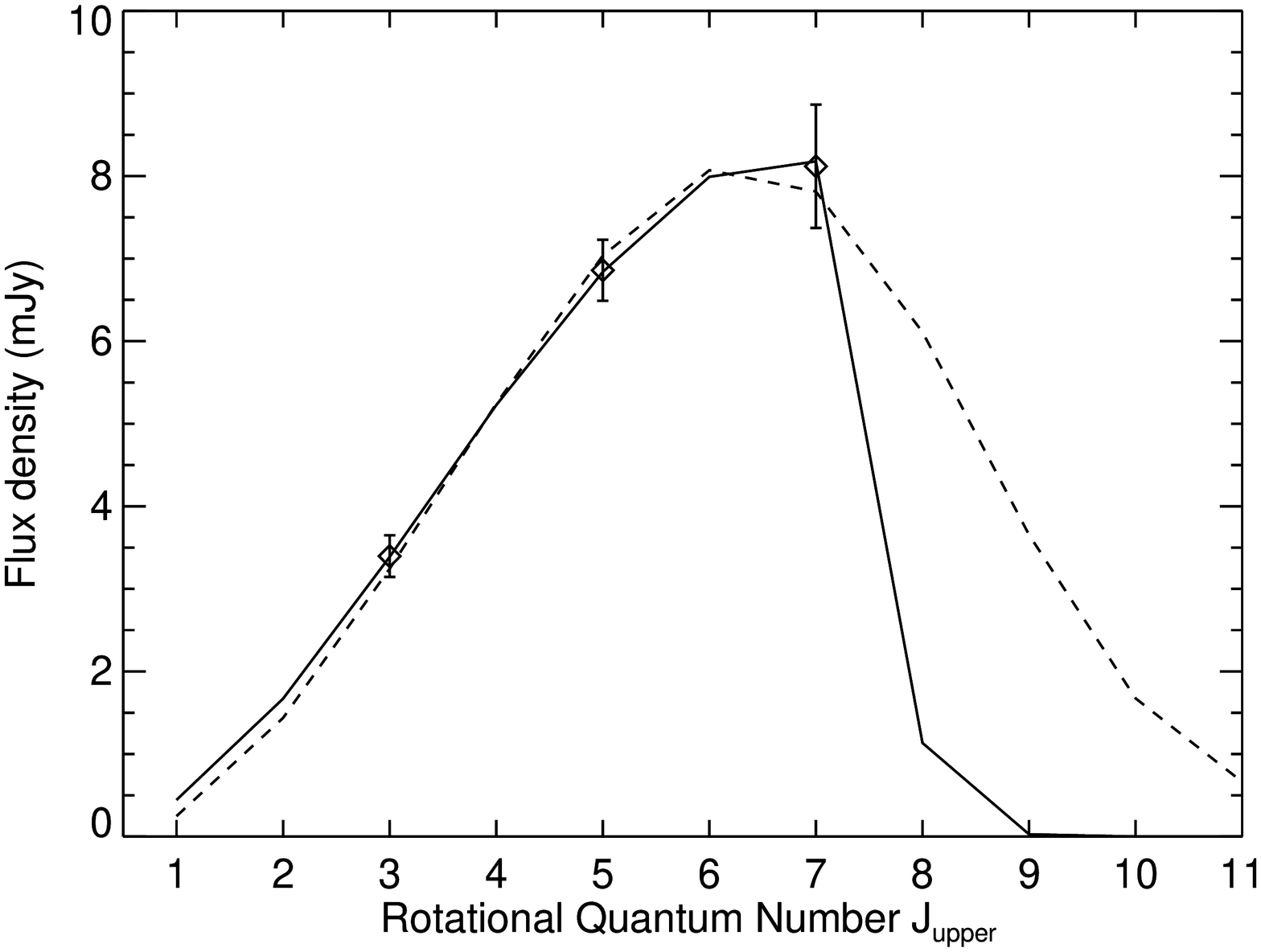, width=6.0cm, angle=90}
\caption{\noindent The CO SED with the SED given by the LVG model giving the lowest $\chi^{2}$ with T$_{\rm{G}}$ = \bestfitT\ K, n(H$_{2}$) = \bestfitN\ cm$^{-3}$, and the CO region size of $r_{0}$ = \bestfitr\ kpc (solid line). This corresponds to the open triangle in Figure \ref{fig:gas_cont}.  A second line, showing a model from within the higher temperature-lower density/size region (T$_{\rm{G}}$ = \regAT\ K, n(H$_{2}$) = \regAN\ cm$^{-3}$ and $r_{0}$ = \regAr\ kpc), corresponding to the open square in Figure \ref{fig:gas_cont}, is there for comparison (dashed line).}
\label{fig:COSED_bestfit}
\end{center}
\end{figure}

At higher transitions ($J\geq8$) the line optical depths are $<1$ and the system is no longer in LTE. The lines undergo collisional de-excitation and the intensities quickly decline to levels no longer detectable above the background radiation. This manifests itself in the sharp decline of the solid line in Figure \ref{fig:COSED_bestfit} above $J=7$. 

The dashed line in this figure is the CO SED corresponding to a secondary minimum $\chi^{2}$ region within the 1$\sigma$ contours of Figure \ref{fig:gas_cont}, with T$_{\rm{G}}$ = \regAT\ K, n(H$_{2}$) = \regAN\ cm$^{-3}$ and $r_{0}$ = \regAr\ kpc (the square in Figure \ref{fig:gas_cont}). Here both the radiative and collisional excitation and de-excitation have significant effects on the line intensities and the full statistical equilibrium analysis must be considered. 

Figure \ref{fig:COSED_bestfit} illustrates the difference between this high-temperature/lower-density solution (the dashed line), and the previous low-temperature/high-density solution (solid line) is most pronounced at the higher $J$ levels. Observations of the $J\geq8$ transitions are needed to further constrain the LVG modelling at these higher transitions.

High resolution imaging of the gas (and dust) in AMS12 would provide a measurement on the size and would constrain the model further. 

\subsubsection{Using the dust temperature PDF as a prior for T$_{G}$}

Here we work under the assumption that the gas and dust arise from the same regions, and are therefore at the same temperature. Indeed dust shields the gas from the ultraviolet and optical radiation preventing it from dissociating the molecules.

Gas and dust studies in other high redshift galaxies have shown both the dust and gas are compact on scales of less than $\sim 4-8$ kpc supporting the assumption that they arise from the same region \citep{Tacconi2006, Riechers2006, Weiss2007, Younger2008}. The dust temperature is often used to either constrain or justify a particular gas kinetic temperature \citep[see for example][ among others]{Weiss2007, Ao2008, Greve2009}.

\begin{figure*}
\begin{center}
\subfigure[]{
\psfig{file = 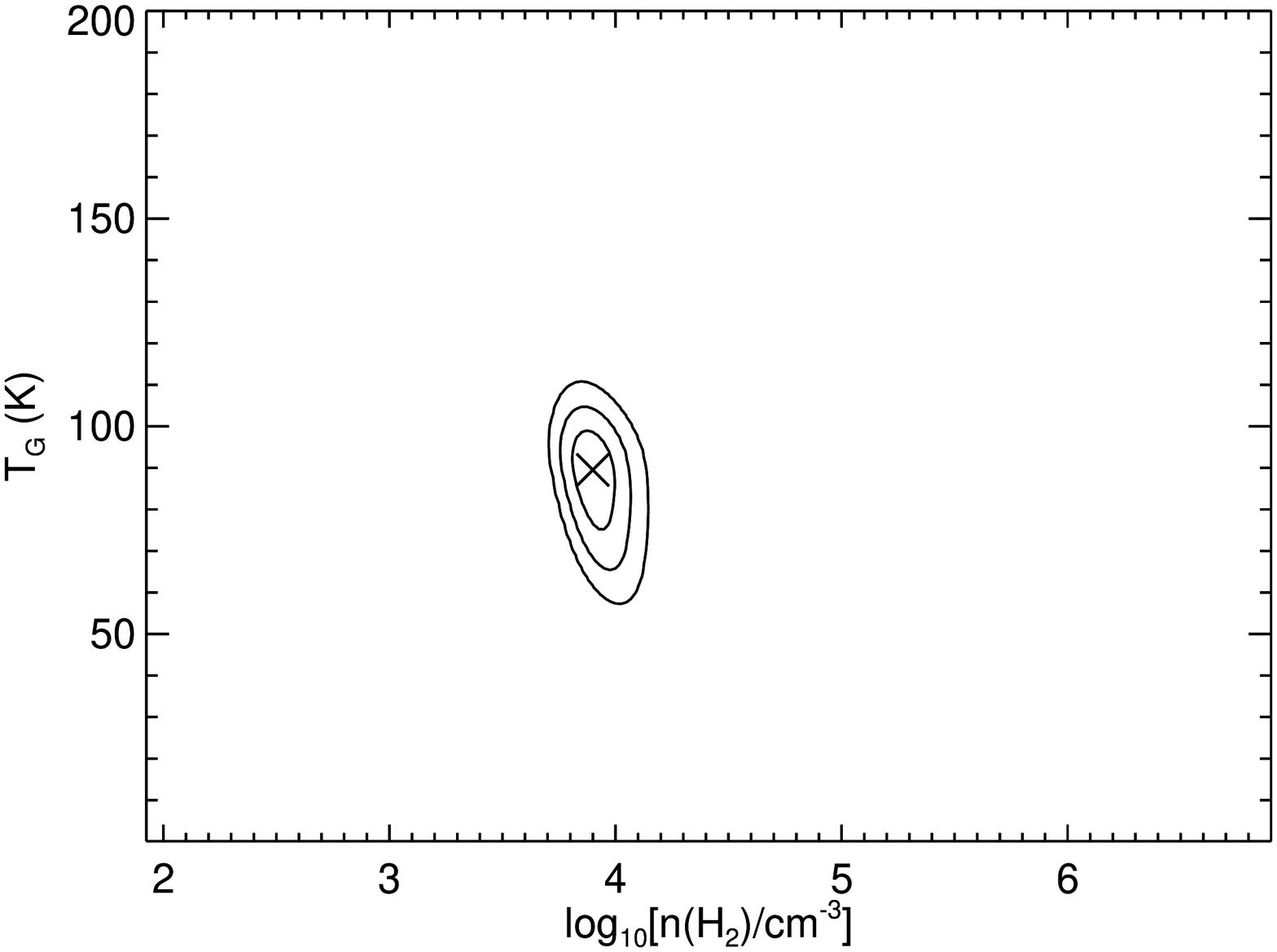, width=5.5cm, angle=90}
\label{fig:prior_cont_TN}
}
\subfigure[]{
\psfig{file=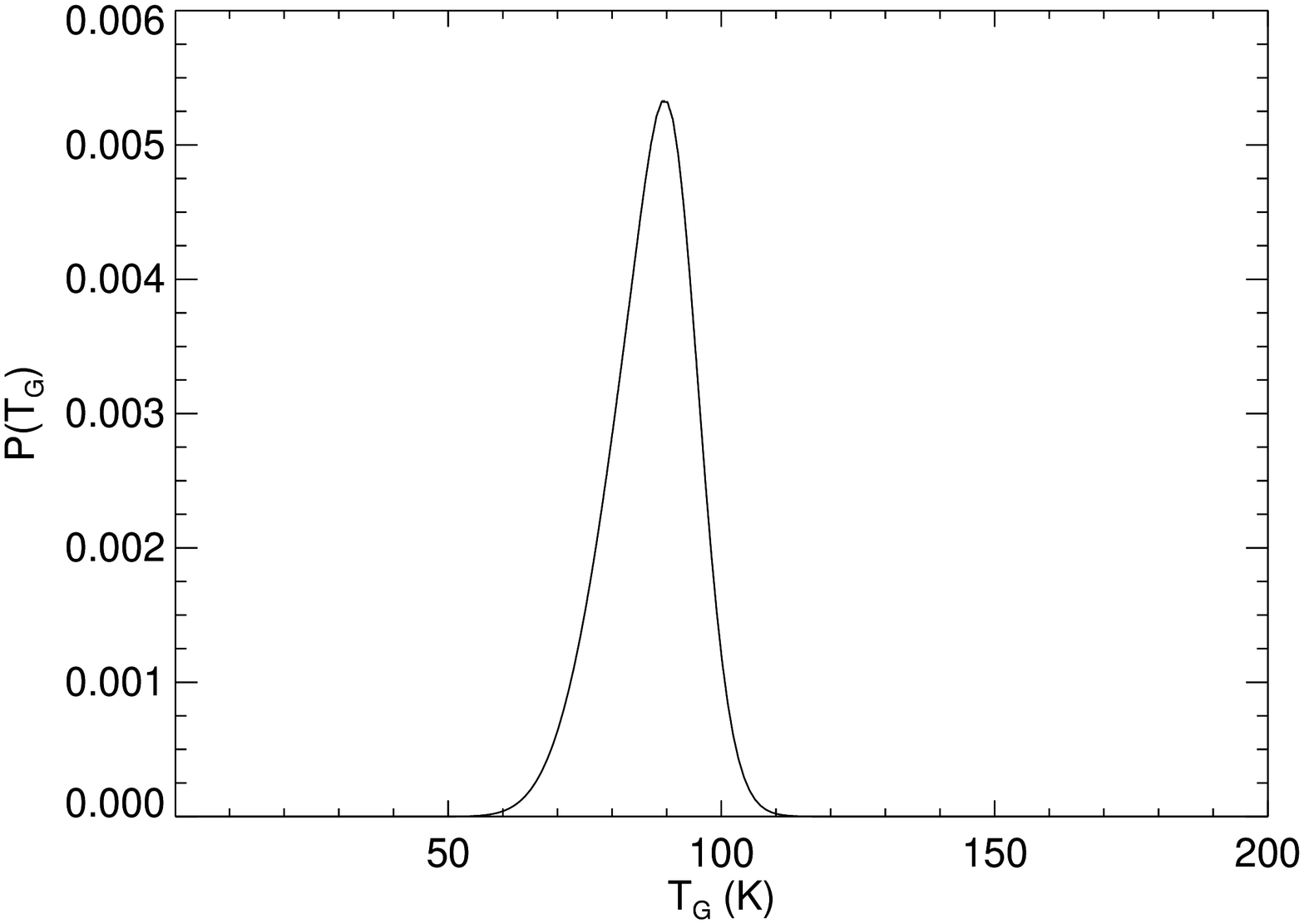, width=5.5cm, angle=90}
\label{fig:marg_prior_T}
}
\subfigure[]{
 \psfig{file=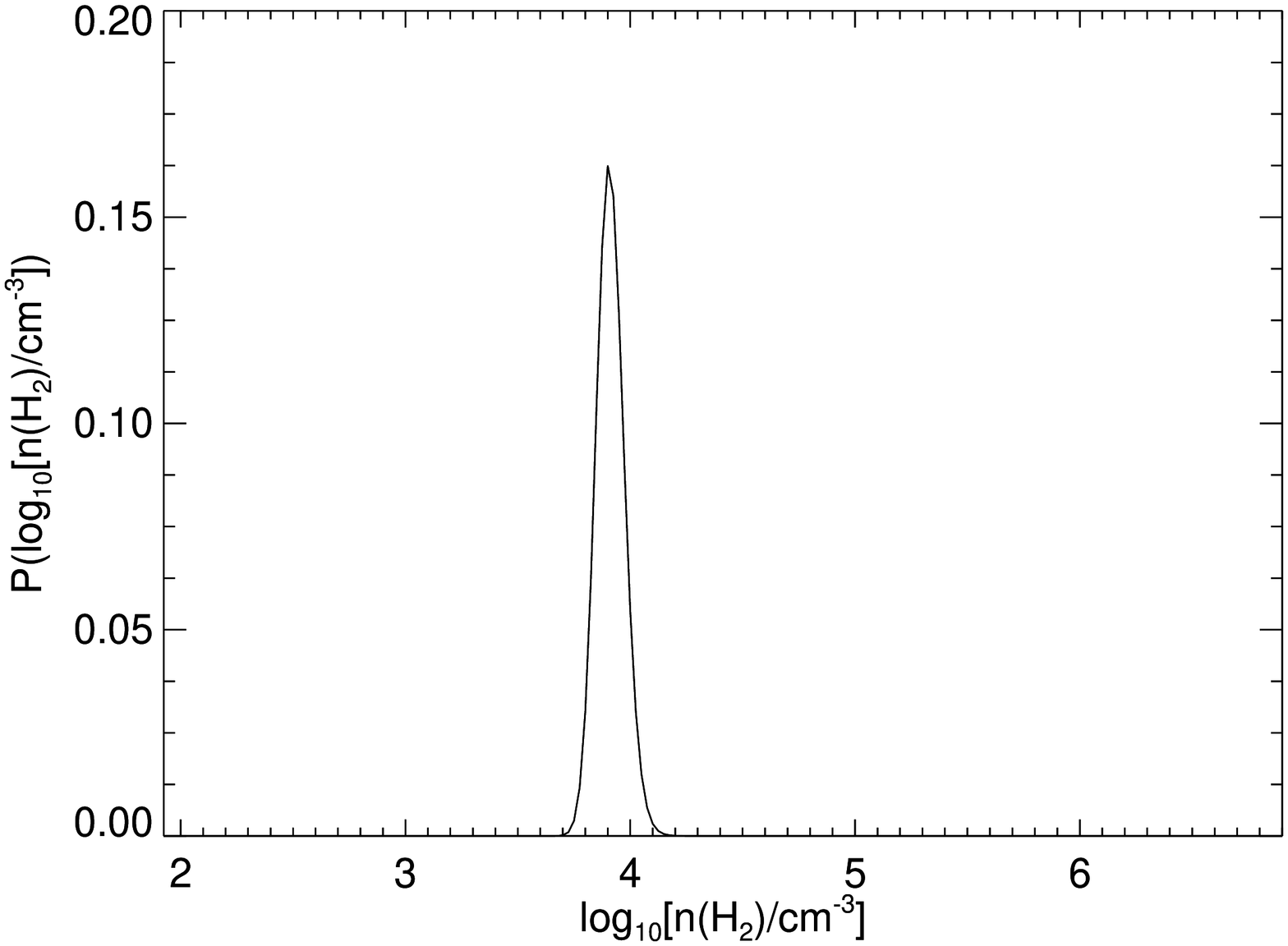, width=5.5cm, angle=90}
 \label{fig:marg_prior_N}
}
\subfigure[]{
\psfig{file = 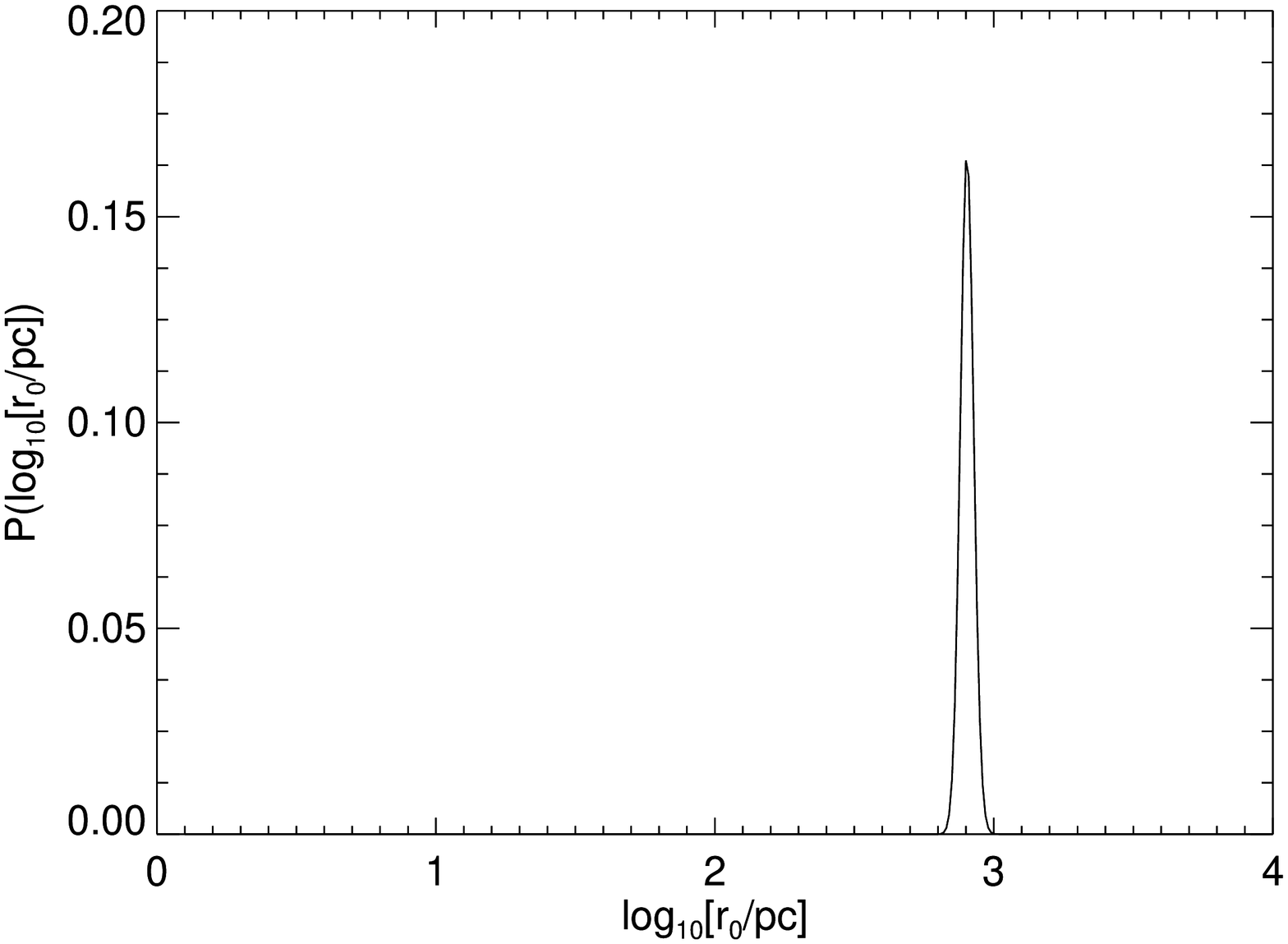, width=5.5cm, angle=90}
\label{fig:marg_prior_r0}
}
\caption{\noindent \textbf{(a):} The 1, 2, and 3$\sigma$ contours of the temperature and density PDFs marginalised over the size. The dust temperature is used as a prior for the gas temperature. The best values of marginalised temperature,  and density are shown in each figure marked with an `X'. \textbf{(b):} The marginalised PDF of the gas temperature using the prior of P(T$_{\rm{D}}$)=P(T$_{\rm{G}}$). Best fit with T$_{G} = 90\pm 8$ K. \textbf{(c):} The marginalised PDF of the gas density parameter log$_{10}$[n(H$_{2}$)/cm$^{-3}$]=3.9$\pm 0.1$.  \textbf{(d):} The marginalised PDF of the size of the emitting region of gas best fit with log$_{10}$[$r_{0}$/pc]=2.9$\pm0.02$. }
\label{fig:cont+prior}
\end{center}
\end{figure*}

Using the PDF of the dust temperature as the prior for the gas temperature, (i.e Figure \ref{fig:dust_temp}, where T$_{\rm{D}}$ peaks at 89 K), P(T$_{\rm{G}}$) = P(T$_{\rm{D}}$) in Equation \ref{eqn:bayes}, we can rule out the lower temperature regions. With the prior on the temperature distribution, the contour map in Figure \ref{fig:prior_cont_TN} shows a tight convergence on the temperature and density. The values of the least $\chi^2$ in this model are approximately equivalent to the peaks of the marginalised PDFs of the parameters which are shown in Figure \ref{fig:cont+prior}. Due to the small number of data points and the shapes of the likelihood and prior PDF, the prior is having a greater effect than the likelihood on the PDF.  The values for the kinetic temperature, density of the gas and the size of the emission region are \margpriorT\ $\pm8$ K, $10^{3.9\pm0.1}$ cm$^{-3}$, and $0.8\pm0.04$ kpc, respectively. 

\begin{figure}
\begin{center}
\psfig{file=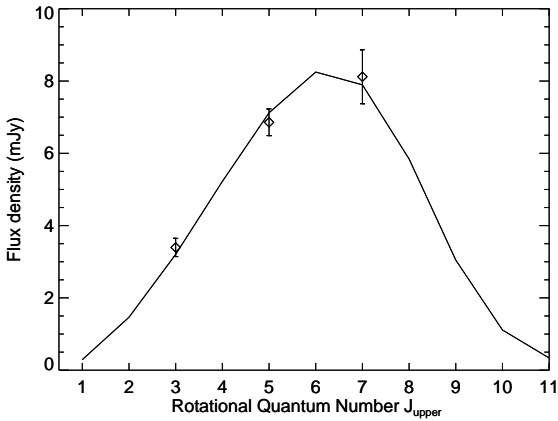, width=6.0cm, angle=90}
\caption{\noindent The CO SED with the SED given by the LVG model with T$_{\rm{G}}$ = \margpriorT\ K, n(H$_{2}$) = \margpriorN\ cm$^{-3}$, and the CO region size of r$_{0}$ = \margpriorr\ kpc, these have the highest probabilities once the prior on the temperature, P(T$_{\rm{D}}$), is applied.}
\label{fig:COSED_prior}
\end{center}
\end{figure}

The LVG model corresponding to these values of temperature, density and emitting region as determined with the use of the dust temperature PDF as a prior for T$_{\rm{G}}$, gives the CO ladder displayed in Figure \ref{fig:COSED_prior}. The combination of these parameter values fall within the 1$\sigma$ contours of Figure \ref{fig:gas_cont}, making them an acceptable fit to the CO SED assuming no prior knowledge.  

\section{Atomic carbon}

The \highCI\ line was detected in AMS12 to an $7\sigma$ significance which prompted a search for the \lowCI\ line. The \lowCI\ line was detected to an $8\sigma$ significance, these lines can be seen in Figure \ref{fig:123mm} and the line properties are given in Table \ref{tab:CO}. This is the first unlensed, high redshift galaxy with both the \lowCI\ and \highCI\ lines detected.

With both the upper and lower fine structure atomic carbon lines detected, we may directly determine physical properties of [CI] in AMS12. 

\subsection{Excitation temperature and mass of atomic carbon}

The excitation temperature of [CI] can be directly determined, assuming the lines are optically thin, from the ratio of L$^{\rm{T}}_{\rm{[CI](^{3}P_{2}-^{3}P_{1})}}$ to L$^{\rm{T}}_{\rm{[CI](^{3}P_{1}-^{3}P_{0})}}$ \citep[e.g.][]{Stutzki1997}. In order to relate the excitation temperature of [CI] to the gas kinetic temperature, we require the [CI] excitation to be in LTE (much in the same way we can estimate the T$_{\rm{G}}$ directly from CO in LTE). If this is not the case, the excitation temperature of [CI] could be lower than the kinetic temperature also.  

The line column densities may be found using the integrated brightness temperatures of the lines \citep[see Appendix A of][for a complete derivation]{Schneider2003}. The excitation temperature is given by the ratio of column densities expressed by the ratio of the statistical weights of the levels and the Boltzmann factor, 

\begin{equation}
\frac{N_{21}}{N_{10}} = \frac{g_{21}}{g_{10}}e^{-\frac{h\nu_{21}}{k{\rm{T}_{ex}}}}
\end{equation}

\noindent rearranging and equating constants the expression for the excitation temperature is found to be,

\begin{equation}
 \rm{T}_{ex} = \frac{38.8}{ln(\frac{2.11}{\textit{R}_{\rm{CI}}})} 
\label{eqn:TCI}
\end{equation}

\noindent where $R_{\rm{CI}}$ is the ratio of the line brightness temperatures, these may be replaced with the L$^{\rm{T}}$ of the lines as L$^{\rm{T}} \propto T_{b}$, $R_{CI}\equiv\rm{L^{\rm{T}}_{\rm{[CI](^{3}P_{2}-^{3}P_{1})}}}/ \rm{L^{\rm{T}}_{\rm{[CI](^{3}P_{1}-^{3}P_{0})}}}$. 

The T$_{\rm{ex}}$ can then be used to find the [CI] total column density and mass. \citet{Weiss2003} derive the beam averaged [CI] column density in the optically thin limit and use this and the area of the emitting region (given by the solid angle subtended by the source convolved with the beam multiplied by the angular distance squared, $\Omega_{s\ast b}D_{A}^{2}$), to derive the mass of [CI]. \citet{Solomon1992} express the line brightness luminosity related to the emitting area, L$^{\rm{T}}=23.5\Omega_{s\ast b}D_{L}I_{CO}(1+z)^{-3}$, where the luminosity distance is related to the angular distance via $D_{A} = D_{L}/(1+z)^{2}$. Thus, \citet{Weiss2003,Weiss2005}, use this expression of L$^{\rm{T}}$  and Equation 3 in \citet{Solomon1992} in order to determine the mass of [CI] in an unresolved source using $\rm{L^{\rm{T}}_{\rm{[CI](^{3}P_{1}-^{3}P_{0})}}}$ via, 

\begin{multline}
\left(\frac{M_{\rm{CI}}}{\rm{M}_{\odot}}\right)= \\
5.706 \times10^{-4}Q(\rm{T_{ex}})\frac{1}{3}\rm{e}^{T_1/T_{ex}}\left(\frac{ L^{T}_{\rm{CI(^{3}P_{1} - ^{3}P_{0}})}}{\rm{K~kms^{-1}~pc^{2}}}\right) 
\label{eqn:MCI}
\end{multline}
 
\noindent where $Q(\rm{T_{ex}}) = 1+3e^{-T_{1}/T_{ex}}+5e^{-T_{2}/T_{ex}}$ is the partition function for [CI]. T$_{1} = 23.6$ K and T$_{2}=62.5$ K are the energies above the ground state. 

\subsection{The [CI] temperature and mass towards AMS12}

The relationships defined above require the [CI] lines to be optically thin. We can test this requirement assuming a T$_{\rm{ex}}$ and size \citep[see equations A6 and A7 in the appendix of][]{Schneider2003}. Firstly, assuming T$_{\rm{ex}}= \rm{T_{G}}$ and a size of 0.8 kpc (i.e. from the LVG model) we infer line optical depths of $\sim0.3$ for both the \lowCI\ and \highCI\ lines. If we assume a larger size (i.e. 2 kpc), the optical depths are $\sim 0.05$. 

From the line luminosities of the atomic carbon in AMS12 (see Table \ref{tab:CO}), we determine the line ratio $R_{\rm{CI}} = 0.85\pm0.24$ and hence a [CI] excitation temperature of $42.7\pm 10$ K. This yields a [CI] mass of $(1.5\pm0.5)\times10^{7}$ \msol. 

The [CI] excitation temperature determined does not fall within either 1$\sigma$ temperature regions in Figure \ref{fig:gas_cont}. Thus when we fit the LVG model with no prior on the temperature, the T$_{\rm{ex}}$ of [CI] is not in agreement with the gas temperatures indicated by high-excitation CO. It is in even less agreement with the gas kinetic temperature we determine using the dust temperature as a prior. This could be either suggesting that the [CI] is not in LTE. Or alternatively, the [CI] emission arises from a more spatially-extended, cooler molecular gas component than the gas probed by the higher-excitation CO lines. 

The optical depth of the [CI] increases with decreasing temperature, using T$_{ex}\approx43$ K the line optical depths are calculated to be $\sim1$ when assuming the compact size of 0.8 kpc. However, assuming the size is instead more extended for example, $2$ kpc, the line optical depths are $\sim0.1$. It is likely that our assumption of optically-thin lines is appropriate as even with the more compact size we are at the limit of optically-thin lines, and the possibility that the [CI] emission region is more extended would lower the line optical depths.   

The LVG model analysis of the high-excitation gas has indicated this to be relatively compact with a radius $\sim1$ kpc. The low-excitation temperature of [CI] which in other high redshift sources is broadly in agreement with the dust temperature \citep{Walter2011}, could be alluding to a second, cooler, more extended region of gas. Since we expect the CO(1-0) to trace the same region as the [CI] emission, if we were to observe CO(1-0) we could test if this low-excitation gas component exists in AMS12. 

This has been seen in high redshift submillimetre galaxies (SMGs). High resolution imaging of CO(1-0) has revealed the CO(1-0) to be more extended than the higher $J$ emission (i.e. CO(3-2) or (4-3)), \citep{Ivison2010, Ivison2011, Carilli2010,Riechers2011b}. While resolved CO(1-0) in strongly lensed quasar hosts shows the CO(1-0) to be compact and similar to the CO(3-2) emission, assuming the low-excitation gas has the same magnification factor as the high-excitation gas \citep{Riechers2006, Riechers2011a}.

\citet{Gerin2000} observed a relationship between L$^{\rm{T}}_{\rm{CO(1-0)}}$ and L$^{\rm{T}}_{\rm{[CI](1-0)}}$ from a survey of low redshift galaxies; L$^{\rm{T}}_{\rm{[CI](1-0)}}$ = 0.2$\pm0.2$ L$^{\rm{T}}_{\rm{CO(1-0)}}$. \citet{Walter2011} observed that their sample of (mostly lensed), high redshift sources support the relationship from \citet{Gerin2000}. For the majority of their sources, CO(3-2) was the lowest transition observed, in order to get an estimate on CO(1-0) they used L$^{\rm{T}}_{\rm{CO(1-0)}}$ = 0.9 L$^{\rm{T}}_{\rm{CO(3-2)}}$. From this, they determined that L$^{\rm{T}}_{\rm{[CI](1-0)}}$ = 0.29$\pm0.13$ L$^{\rm{T}}_{\rm{CO(1-0)}}$.  

However, in the case of one SMG (SMM J163658+4105), which \citet{Ivison2011} observed in CO(1-0), the ratio of the brightness temperature luminosities of CO(3-2) and CO(1-0) is actually $0.54\pm0.12$, significantly lower than the assumed 0.9 that \citet{Walter2011} used. Using the CO(1-0) strength from \citet{Ivison2011}, the ratio of L$^{\rm{T}}_{\rm{CO(1-0)}}$/L$^{\rm{T}}_{\rm{[CI](1-0)}}$ then becomes 0.14, in disagreement with the high redshift relationship from \citet{Walter2011} and closer to the low redshift relationship. The ratio of L$^{\rm{T}}_{\rm{CO(1-0)}}$/L$^{\rm{T}}_{\rm{[CI](1-0)}}$ at high redshift might well be similar to that at low redshift, once differential magnification of high- and low-$J$ CO lines in lensed sources, has been taken into account (e.g. Deane et al. 2012b, in prep.).

\section{Discussion} \label{sec:disc}

\subsection{Gas mass} \label{sec:GasMass}

From the measured brightness temperature luminosity of the CO gas ($L^{\rm{T}}_{\rm CO}$), the $\rm H_{2}$ mass can be found using the relation:

\begin{multline}
\left(\frac{M(\rm H_{2})}{\rm{M_{\odot}}}\right) = \\
\left( \frac{\alpha}{M_{\odot}(\rm{K~kms^{-1}~pc^{2}})^{-1}}\right) \left(\frac{\rm{L^{\rm{T}}_{\rm CO}}}{\rm{K~kms^{-1}~pc^{2}}}\right)
\label{eqn:COH}
\end{multline}

\noindent \citep[see][]{SolVan2005}, where $M(\rm H_{2})$ includes He and is therefore $\approx M_{\rm gas}$. \citet{Downes1998} determined the constant $\alpha$ empirically from a study of a sample of ultra-luminous infrared galaxies (ULIRGs) and high redshift galaxies to be $\alpha~=~0.8$. We assume this value of $\alpha~=~0.8$ hereafter, although we note that the acceptable values are in the range 0.3 - 1.3 \citep[see Table 9 of][]{Downes1998}.

The brightness temperature luminosity of CO is best represented by L$^{\rm{T}}_{\rm{CO(1-0)}}$. This requires a measurement of CO(1-0) that we do not have, but we can calculate estimates of the CO(1-0) flux via various methods. Three of these methods presented here assume that the molecular gas is represented by a single highly-excited component. We can also use the \lowCI\ line and the relationship from \citet{Gerin2000} to estimate the CO(1-0) brightness temperature luminosity without this assumption.  

\subsubsection{Method A}

Assuming the transitions lower than CO(3-2) are thermalised, we can also derive a total gas mass, $M$(H$_{2}$), from L$^{\rm{T}}_{\rm{CO}(1-0)}$ using the relationship in Equation \ref{eqn:COH}.

Under the assumption that CO(3-2) is thermalised, L$^{\rm{T}}_{\rm{CO(3-2)}}$ = L$^{\rm{T}}_{\rm{CO(1-0)}} = (4.2\pm0.4) \times 10^{10}$ K \kms\ pc$^{2}$. This gives a gas mass of $M(\rm{H_2})\sim3.3\times 10^{10}(\frac{\alpha}{0.8}$)\msol.

\subsubsection{Method B} 

Method B uses the best fit LVG model shown in Figure \ref{fig:COSED_bestfit} (solid line, T$_{\rm{G}}$ = \bestfitT\ K, n(H$_{2}$) = \bestfitN\ cm$^{-3}$). The CO(1-0) line brightness temperature T$_{\rm{b}}$ is given by the LVG model, and following the use of Equation \ref{eqn:sco} to convert to observed flux, we calculate L$^{\rm{T}}_{\rm{CO(1-0)}} \approx 4.8\times 10^{10}$ K \kms\ pc$^2$. The gas mass determined from Method B is $M(\rm{H_2}) \sim 3.9 \times 10^{10}(\frac{\alpha}{0.8}$)\msol. 

\subsubsection{Method C}

We have used the dust information we have for this source and applied the prior distribution for the gas temperature P(T$_{\rm{G}}$) = P(T$_{\rm{D}}$). The best fit model is shown in Figure \ref{fig:COSED_prior} (T$_{\rm{G}}$ = \margpriorT\ K, n(H$_{2}$) = \margpriorN\ cm$^{-3}$). We can again use this LVG model's CO(1-0) T$_{\rm{b}}$ and convert it to a luminosity of L$^{\rm{T}}_{\rm{CO(1-0)}} \approx 3.2\times 10^{10}$ K \kms\ pc$^{2}$. This gives $M_{\rm{H_{2}}} \sim 2.6 \times 10^{10}(\frac{\alpha}{0.8}$)\msol.

\subsubsection{Method D}

Using the relationship relating L$^{\rm{T}}_{\rm{CO(1-0)}}$ to L$^{\rm{T}}_{\rm{[CI](1-0)}}$ from \citet{Gerin2000}, we estimate the L$^{\rm{T}}_{\rm{CO(1-0)}} = 6.1\times10^{10}$ K \kms\ pc$^{2}$. We use the low redshift relationship as the results from \citet{Walter2011} are estimated from the higher CO(3-2) transition and not CO(1-0). The gas mass inferred from this CO(1-0) strength is $M_{\rm{H_{2}}} \sim 4.9 \times 10^{10}(\frac{\alpha}{0.8}$)\msol. This is larger than the masses determined from the previous three methods which is expected if the [CI] observations are revealing a low-excitation, more diffuse region of gas, i.e. such as the low-excitation gas reservoirs seen in the CO(1-0) observations of some SMGs \citep{Ivison2010,Ivison2011,Carilli2010,Riechers2011b}. 

\bigskip
The results are displayed in Table \ref{tab:CI_temp}. The first three methods (\textit{A} to \textit{C}), assume that the high-$J$ CO lines trace the total molecular gas. The variance between the mass estimates from these three methods yields an estimate of the uncertainty in the mean value $\langle M_{\rm{H_{2}}}\rangle=(3.3\pm0.9)\times10^{10}(\frac{\alpha}{0.8})$ \msol. However, with both [CI] lines detected we have found a [CI] excitation temperature which does not agree with either of the LVG model solutions. It is thought that [CI] traces the low-excitation CO, as the critical densities of [CI] and CO(1-0) are very similar and there is observational evidence of [CI] and CO arising from the same regions \citep[e.g.][]{Ikeda2002}. The assumption that the high-excitation CO lines trace CO(1-0) may be incorrect and lead us to underestimate the total gas mass.

Indeed, studies of high redshift SMGs, show the resolved CO(1-0) evidently arise from extended, low-excitation gas \citep[see for example][and references therein]{Ivison2010,Ivison2011,Carilli2010}. \citet{Ivison2011} found that using CO(1-0) to determine the gas mass in four SMGs at $z \grtsim 2$ gave masses $\sim 2$ times higher than masses determined from the CO(3-2) or higher $J$ lines, i.e. the assumption that the higher $J$ transition (CO(3-2) or CO(4-3) in most cases), is thermalised was incorrect. 

\begin{table*}
 \begin{minipage}{160mm}
  \begin{center}
    \begin{tabular}{|c|c|c|c|}
      \hline
      \hline
Method&L$^{\rm{T}}_{\rm{CO(1-0)}}/10^{10}$&$M_{\rm{H_2}}/10^{10}$&L$^{\rm{T}}_{\rm{[CI](1-0)}}$/L$^{\rm{T}}_{\rm{CO(1-0)}}$\\
&$\rm{[K~kms^{-1}~pc^{2}]}$&[\msol]&\\
\hline
A&4.181&3.34& 0.29\\
\hline
B& 4.837&3.87&0.25\\

\hline
C& 3.246&2.60&0.37\\ 
\hline
D& 6.085&4.89&0.20$^{*}$\\
\hline
\hline
    \end{tabular}
  \end{center}
\caption{\noindent The calculated CO(1-0) line strength determined four ways. \textit{A}: from the assumption that CO(3-2) is thermalised; \textit{B}:  from the unconstrained best fitting LVG model; \textit{C}: from the LVG model assuming P(T$_{\rm{D}}$) = P(T$_{\rm{G}}$); and \textit{D}: from the \lowCI\ strength using the \citet{Gerin2000} relationship.  The [CI]/CO relationships for each CO(1-0) are determined and displayed. $^*$ This is the \citet{Gerin2000} relationship that we have assumed to derive the CO(1-0) mass.}
   \label{tab:CI_temp}
 \end{minipage}
 
\end{table*}

Single component LVG models of SMGs have been shown to underestimate the CO(1-0) line \citep[i.e.][]{Carilli2010,Riechers2011b}. Direct observations of the CO(1-0) line in AMS12 are needed to test whether a significant component of low-excitation gas is present. 

Hence, from here on we shall use the gas mass determined from the CO(1-0) strength estimated from the \lowCI\ line (method \textit{D}), $M_{\rm{H_{2}}} \sim 4.9 \times 10^{10}(\frac{\alpha}{0.8}$)\msol, although we acknowledge there is considerable uncertainty in this estimate due to the large scatter in the L$^{\rm{T}}_{\rm{CO~(1-0)}}/\rm{L^{T}_{[CI](1-0)}}$ relationship.

\subsection{Discriminating between LVG models}

We discuss various methods we use to discriminate between the LVG models. First, we consider the models' outputs and the constraints future observations can make, secondly, by estimating the mass given by the models assuming the gas is distributed in a thin disc, we compare these with the gas mass from Section \ref{sec:GasMass}. Then we explore the possibility of a two component LVG model inspired by the excitation temperature of [CI].

\subsubsection{From the LVG models' output line brightness temperatures} 
The sharp decline in the intensity of the higher-$J$ transitions in the low-temperature/high-density model can be used with future observations to discriminate between the models. Using the output T$_{\rm{b}}$'s we can predict the possible line strengths of the higher $J$ transitions, particularly using the CO(8-7) and CO(9-8) transitions which happen to be in observable windows. These would be $S_{\rm{CO(8-7)}} = 1.1$ mJy and $S_{\rm{CO(9-8)}} =  0.03$ mJy respectively (see Figure \ref{fig:COSED_bestfit}).  However, if we instead use the T$_{\rm{b}}$'s from the solution where we used the dust temperature as a prior for the gas temperature the line strengths are $S_{\rm{CO(8-7)}} = 5.9$ mJy and $S_{\rm{CO(9-8)}} = 3.1$ mJy respectively (see Figure \ref{fig:COSED_prior}). 

The significant differences in these line strengths mean that observations of higher CO transitions in this object, would conclusively constrain the region of parameter space of the LVG model which describes the conditions of the gas. 

\subsubsection{From mass estimates using volume arguments}
Another way we may be able to rule out one of the models is by considering the spatial distribution of the gas. Consider the LVG model which gives the lowest $\chi^{2}$ value, T$_{\rm{G}}$ = \bestfitT\ K, n(H$_{2}$) = \bestfitN\ cm$^{-3}$ and $r_0$ = \bestfitr\ kpc, and assume the molecular gas is distributed in a thin disc with height $H$, of uniform density. We can calculate the volume of the gas given the density. We use the findings of \citet{Downes1998} that the molecular gas in the central regions of ULIRGs is not a collection of separate clouds undergoing self-gravitation, but rather clouds fused together to form a disc of more or less constant density. They modelled the structure of these gas discs and found the average height of the discs to be $H\sim 58$ pc. Taking this value as the height of the assumed disc of gas for AMS12, we can calculate the volume of the disc given $r_{0}$ = \bestfitr\ kpc. 

From this volume, and the best fitting density of the gas n(H$_2$) = \bestfitN\ cm$^{-3}$, we can estimate the gas mass in this volume to be $\sim 4.1 \times 10^{14}$ \msol. This is four orders of magnitude higher than the gas mass derived from our observations (see Section \ref{sec:GasMass}). Table \ref{tab:CI_temp} shows the gas mass M$_{\rm{H_2}}$, for the four methods described in Section \ref{sec:GasMass}. 

On the other hand, the LVG model giving the parameters T$_{\rm{G}}$ = \margpriorT\ K, n(H$_{2}$) = \margpriorN\ cm$^{-3}$ and $r_{0}$ = \margpriorr\ kpc, gives a mass of $4.6 \times 10^{10}$ \msol\ assuming the same disc height of 58 pc. This is very similar to the masses in Table \ref{tab:CI_temp} determined by Equation \ref{eqn:COH}. 

Though we note we have made various assumptions to estimate the gas masses in Section \ref{sec:GasMass}, this crude mass estimate from the volume argument, distinguishes the LVG model solutions from each other. The estimate from the volume of the low-temperature/high-density solution does not agree with the gas masses determined in Section \ref{sec:GasMass}, while the mass from the solution using the dust temperature as a prior, does agree. 

\subsubsection{Two-component model}\label{sec:twocomp}
\citet{Riechers2011b} found in the two SMGs they studied, the CO ladders were best fit with two component LVG models; a dense, high-excitation component to fit the higher-$J$ CO lines, and another low-excitation component to fit the low-$J$ CO measurements. 

To investigate the possibility the [CI] gas in AMS12 is from a lower-temperature, more diffuse region, we fit a second component to the LVG model. We used the CO(1-0) estimate from the \lowCI\ line, and for the high-excitation component we keep the previous LVG model solution using the dust prior (from Figure \ref{fig:COSED_prior}).  

\begin{figure}
\begin{center}
\subfigure[]{
\psfig{file = 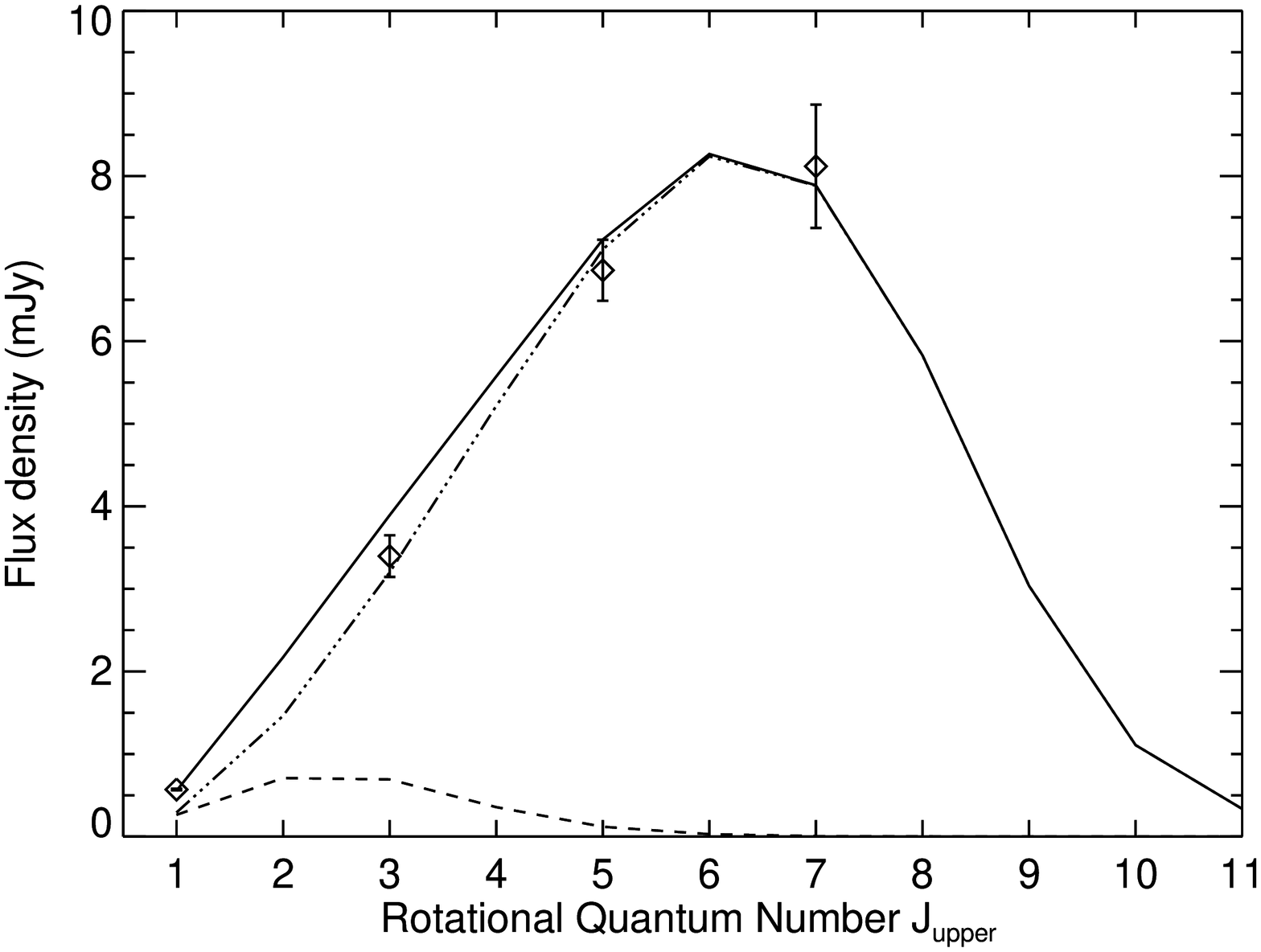, width=5.5cm, angle=90}
\label{fig:2compfixd}
}
\subfigure[]{
\psfig{file=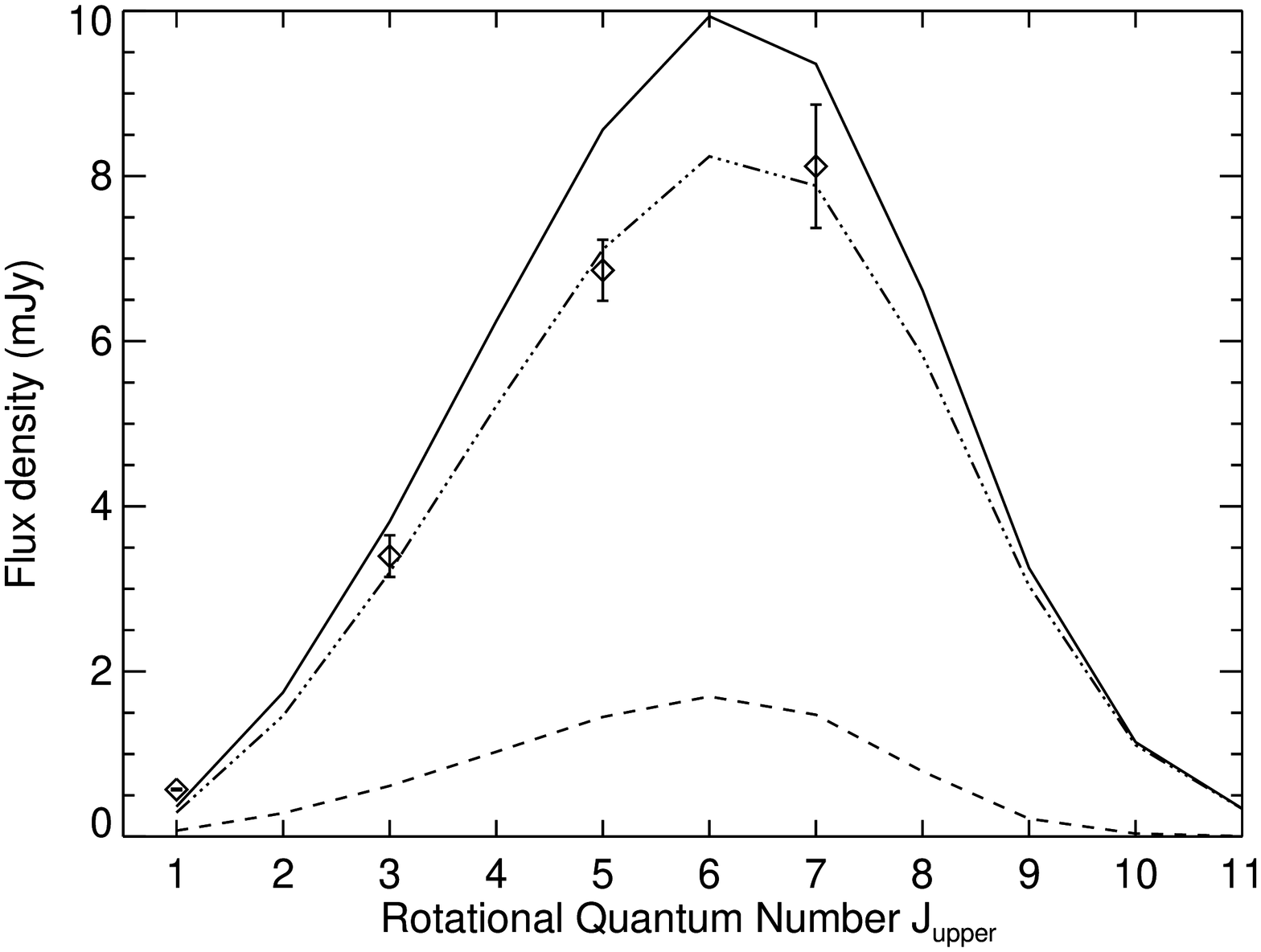, width=5.5cm, angle=90}
\label{fig:2compfixr1}
}
\subfigure[]{
\psfig{file=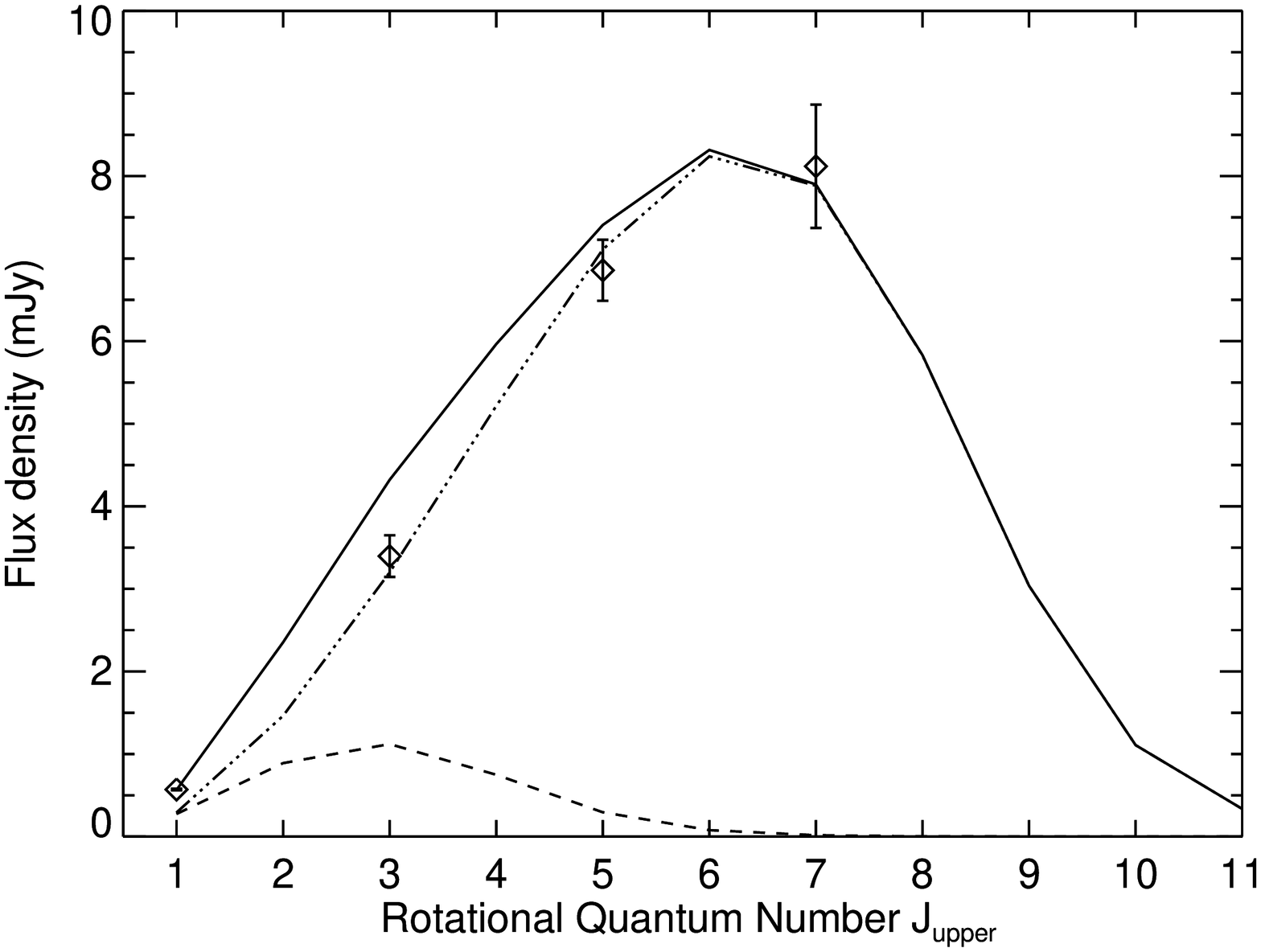, width=5.5cm, angle=90}
\label{fig:2compfixr2}
}
\caption{\noindent Two component LVG models fitted to the CO observations and an estimated CO(1-0) line strength of 0.57 mJy. The low-excitation component is given by the dashed line, while the high-excitation component (from Figure \ref{fig:COSED_prior}) is given by the dot-dot-dash line. The combined model is given by the solid line. \textbf{(a):} The low-excitation component with T$_{\rm{G}}\approx$ 43 K, n(H$_{2})=10^{2.5}$ cm$^{-3}$ and size of $r_0$ = 2.8 kpc. \textbf{(b):} The low-excitation component in this figure has the solution T$_{G} = 42.7$ K, n(H$_{2})=10^{3.4}$ cm$^{-3}$ and a fixed size of $r_{0}$ = 1 kpc. \textbf{(c):} The low-excitation component here has T$_{\rm{G}}\approx$ 43 K, n(H$_{2})=10^{2.1}$ cm$^{-3}$ and size of $r_0$ = 6 kpc.}
\label{fig:2comp}
\end{center}
\end{figure}

We fix the temperature of the second component at 42.7 K (the [CI] excitation temperature), and tie both components to the estimate of the CO(1-0) strength. Since we only have one data point, we cannot fit for both the size and the density at the same time, so we fit each independently keeping the other fixed at a fiducial value. Firstly, we keep the density fixed to $\approx 10^{2.5}$ cm$^{-3}$ and fit for the size. This density is chosen as a representative of a diffuse component and corresponds to the critical density of CO(1-0). The best fit to the radius is 2.8 kpc and the resulting LVG model is shown in Figure \ref{fig:2compfixd}.

Next, we fix the size of the low-excitation gas, while continuing to hold the temperature at 42.7 K, and fit for the density. We choose two arbitrary sizes; 1 kpc, which is similar to the best fitting size of the single component LVG model solution to the high-$J$ CO lines, and 6 kpc, which is significantly more extended. For the second component solution with a fixed size of 1 kpc, the best fitting density is n(H$_2$) = 10$^{3.4}$ cm$^{-3}$. The two component LVG model (comprising of this solution and the best fitting solution of the high-$J$ CO lines with the dust temperature prior applied), is shown as the solid line in Figure \ref{fig:2compfixr1}. It does not provide a good fit to the observations. 

With the second component size fixed at 6 kpc, (and temperature remaining at 42.7 K), the best fitting density is n(H$_{2}$) = 10$^{2.1}$ cm$^{-3}$. The corresponding two component LVG model is shown in Figure \ref{fig:2compfixr2}. This second component solution provides a better fit to the observed CO lines. The estimates shown in Figures \ref{fig:2compfixd} and \ref{fig:2compfixr2}, are consistent with a more diffuse and extended gas which is only detectable via the low-$J$ CO lines and the [CI] lines, while still providing good fits to the high-$J$ observations. 

We note that by construction, our method will only select solutions with weak emission of the high-$J$ lines. Since we have begun by fitting CO(3-2), CO(5-4) and CO(7-6) with a single component model, the extra component used to fit CO(1-0) must produce negligible flux in the CO(3-2), CO(5-4) and CO(7-6) lines, otherwise their predicted fluxes will be higher than those observed.  

\subsection{Dynamical mass}

We can estimate the dynamical mass of the system from the FWHM of the CO lines, assuming a characteristic radius via \citep{Neri2003};

\begin{equation}
\left(\frac{M_{\rm{dyn}}\rm{sin}^{2}\textit{i}}{\rm{M_{\odot}}}\right) = 4\times10^{4}\left(\frac{\Delta{\rm{V_{FWHM}}}}{\rm{km s^{-1}}}\right)^{2}\left(\frac{r}{\rm{kpc}}\right)
\end{equation}

Studies have shown the virial mass estimate is reasonable even if the gas is clumpy \citep[see][]{Daddi2010}. 

We can use the CO linewidths to estimate the dynamical mass, however, if the [CI] traces a low-excitation region of gas, we should use the [CI] linewidths. Given the large uncertainties in the linewidths in Table \ref{tab:CO}, the CO and [CI] linewidths are very similar and agree within 2$\sigma$. Using both the [CI] linewidths, we use $\Delta\rm{V_{FWHM}} \approx260$ \kms. This will give an approximate estimate on the dynamical mass. Assuming two arbitrary radii encompassing a range of sizes, i.e. the radii from Section \ref{sec:twocomp}, $r_{1} = 1$ kpc and $r_{2}=6$ kpc, we can estimate two values of the dynamical mass of AMS12. Using $r_1$, the dynamical mass is $M_{\rm{dyn}}\rm{sin}^{2}\textit{i} \approx 2.7\times10^{9}$ \msol, while using $r_2$, gives $M_{\rm{dyn}}\rm{sin}^{2}\textit{i} \approx 1.6\times10^{10}$ \msol.

The dynamics are dominated by the molecular gas and stars, with the dark matter and ionized hydrogen sub-dominant \citep{Daddi2010}. The stellar mass in AMS12 is found from the mid-/near-infrared SED by \citet{Lacy2011}, to be $M_{\star}\approx3\times10^{11}$ \msol. The dynamical mass estimates make it difficult to account for the stellar mass. It may be that the radius estimates we have used are not adequate to encompass all the stellar mass.  The constraints on the inclination angle assuming the radii above are severe when considering the stellar and gas mass, for example, if $r_{1}=1$ kpc, then the inclination angle \textit{i}$\lesssim5\degree$. While using the radius estimate $r_2=6$ kpc, the constraint on the inclination angle relaxes slightly to $\textit{i}\lesssim13\degree$. These arguments suggest the host galaxy is seen face-on. 

The radio spectrum of AMS12 has been investigated in \citet{Martinez2006b,Klockner2009}. This object has a steep extended radio spectrum, and narrow optical emission lines pointing to torus obscuration, i.e. the orientation of the central engine and its obscuring material is closer to edge-on to the observer. The low inclination angle of the host from the dynamical mass estimates, suggest the central AGN region and the galaxy's stellar, gas and dust regions are not aligned. Together with the observed narrow emission lines, this suggests AMS12 is obscured by the torus and not by dust in the host galaxy \citep[see][]{Martinez2006a}.

The caveats to using this dynamical mass estimate are many. The high-excitation CO lines may be tracing a separate gas component than the [CI] lines, and therefore the radius estimates based on the high-excitation CO LVG modelling are tenuous for the possible low-excitation component. Though the [CI] linewidths are similar to the CO linewidths, this does not immediately place the [CI] at the same region of the high-excitation CO, i.e. if [CI] is alluding to a more massive while more extended low-excitation gas reservoir. Resolving the detected CO and [CI] lines would constrain the sizes of these emitting regions, which may be possible with the upcoming IRAM Northern Extended Millimeter Array (NoEMA) upgrade to PdBI. In particular though, resolved CO(1-0) in this object would significantly improve the dynamical mass estimate in this object.

\subsection{[CI] abundance and cooling contribution in AMS12}

The abundance of [CI] in AMS12 can be determined by $X[\rm{CI}]/\textit{X}[\rm{H_2}] = \textit{M}_{\rm{CI}}/(6\textit{M}_{\rm{H_{2}}})$, where $M_{\rm{[CI]}}\sim1.5\times10^{7}$ \msol. The abundance assuming the high-excitation gas can be used to estimate the gas mass is $7.8\times 10^{-5}$. While if there is a lower temperature, more diffuse component of the gas, the M$_{\rm{H_{2}}}$ would be higher. The M$_{\rm{H_{2}}}$ estimated from the \lowCI\ line is $\approx5\times10^{10}$ \msol, giving a [CI] abundance of  $X[\rm{CI}]/\textit{X}[\rm{H_2}] = 5.2\times10^{-5}$. This indicates the molecular gas is already enriched at this redshift, supporting findings from \citet{Walter2011}. 

The ratio of L$_{\rm{[CI](1-0)}}$/L$_{\rm{FIR}}$ provides a measure of the cooling contribution of [CI]. For AMS12, L$_{\rm{[CI](1-0)}}$/L$_{\rm{FIR}}=1.5\times10^{-6}$, which appears to be typical for the quasar sources of \citet{Walter2011}. Their sources are split into quasars and SMGs, and while quasars have L$_{\rm{[CI](1-0)}}$/L$_{\rm{FIR}}$ ratios similar to that of AMS12, the SMGs ratios are around an order of magnitude higher. This could be due to the AGN contribution to the L$_{\rm{FIR}}$ in the quasars, as discussed earlier. Overall, the [CI] lines are not major coolants, in fact they are negligible compared to the cooling by the dust continuum.

\subsection{AGN bolometric luminosity and the scale of AGN heating}

From our current observations of the dust and high-excitation gas in AMS12 the temperatures we determine are shown to be beyond the capabilities of heating by star formation alone. Empirical observational evidence and radiative models of star-forming galaxies show that the dust temperatures reach $\lesssim50$ K \citep[see for example][]{Kovacs2006,Siebenmorgen2007}. The typical dust temperatures derived from FIR SED fitting in studies of high redshift star-forming galaxies, range from 30-60 K with the average being $\sim35$ K \citep[e.g.][]{Kovacs2006,Coppin2008, Elbaz2011}. These temperatures are typical of local galaxies where heating of the dust is dominated by young stars \citep[e.g.][]{Farrah2003, Elbaz2011}.  

However, in the hosts of luminous AGN, the FIR emission could also be heated by the AGN. APM 08729+5255, F10214, BR 1202-0725 and Cloverleaf for example, have hotter FIR dust than typical star-forming galaxies and other SMGs. Detailed studies of the dust and gas in these objects have revealed the dust is compact which supports the possibility of significant AGN heating \citep{Solomon2003, Riechers2006, Weiss2007, Ao2008}. 

We can estimate the scale of the heating from the AGN using the bolometric luminosity of AMS12 from the broad-band data between 3.6 and 24 $\mu$m \citep[as done by][]{Martinez2009}. For AMS12, an AGN of \lbol\ = $2\times10^{13}$ \lsol, we assume that L$_{uv}$ is $\sim$ 0.25\lbol, and using the FIR SED fitted parameters ($T_{D}~=~88$ K and $\beta~=~0.6$), we find the scale of AGN heated dust out to 88 K is 2.7 kpc. This result is obtained assuming that the ultraviolet photons travel unhindered up until the radius where the dust becomes self-shielding (i.e. at large optical depths), thus 2.7 kpc may be thought of as the characteristic radius to which the dust is heated to this temperature by the AGN \citep[see][for details]{Barvainis1987}. Clearly, since the characteristic scale of the dust is indeed found to be $\sim2$ kpc in many objects \citep{Greve2005,Tacconi2006,Younger2008}, the dust temperature observed in AMS12, T$_{\rm{D}}$ = 88 K, could be achieved through heating from the AGN. 
 
Note, this estimate of the bolometric luminosity does not take the \lfir\ into account, and therefore may be considered a lower limit. If we account for the \lfir\ which we believe is also attributed to the AGN ($3\times10^{13}$ \lsol), our new estimate of the \lbol\ is $\approx5\times10^{13}$ \lsol\ for AMS12, which could heat the dust to 88 K out to around $\sim4$ kpc. 

\subsection{Black hole, stellar, gas and dust masses}

We estimate the dust mass to be between $9.2\times10^{7}$ \msol\ and $1.6\times10^{9}$ \msol\ using the \lfir\ and two different mass absorption coefficients (this range is typical for high redshift sources, see for example \citet{SolVan2005}). The dust masses vary greatly due to the process of extrapolating the mass absorption coefficients to the rest frequencies, a power law which depends upon $\beta$. The gas mass from the [CI] observations is determined to be $4.9\times 10^{10}$ \msol. This yields a gas-to-dust mass ratio between $\sim30-530$.

We can estimate the Eddington limited black hole mass of this system given the \lbol. If accreting at $\lesssim100\%$ of the Eddington rate \citep[reasonable for quasars at high redshifts, see][]{Mclure2004}, the black hole mass estimated from the \lbol\ determined from the mid-infrared SED is $M_{\bullet}\grtsim6\times10^{8}$ \msol. The revised \lbol\ (including the \lfir), leads to an estimate of the black hole mass $M_{\bullet} \grtsim 1.5 \times10^{9}$ \msol. We have assumed that all of the \lfir\ is attributed to heating from the AGN, though we acknowledge that star formation is likely to contribute, we do not have a measure of the extent of this contribution.  
 
The stellar mass of AMS12 has been estimated by \citet{Lacy2011} to be $M_{\star} = (3.2\pm0.3)\times10^{11}$ \msol. We assume this stellar mass is located entirely within the bulge as AMS12 is probably the progenitor of a present-day elliptical galaxy. This means AMS12 has already the stellar mass of a present-day 2L$^{*}$ galaxy. 

Given the gas mass of $M_{\rm{H_{2}}} \approx 4.9\times10^{10}$ \lsol\ we derive from the CO observations, were this to be converted into stars with 100\% efficiency, it would still only increase the stellar mass by $\sim15\%$. Observations of CO(1-0) are needed to probe the lower-excitation gas and give a more accurate value of the gas mass. However, it seems unlikely that the stellar mass will increase significantly, unless we have underestimated the gas mass by a factor of $\sim10$. 

Assuming $M_{\star} = M_{bulge}$, we have a $M_{\bullet}/M_{bulge}$ ratio for AMS12 of $\grtsim 0.005$. This is significantly higher than the relationship determined from nearby ($z\sim0$) galaxies by \citet{Marconi2003} and \citet{Haring2004} where $M_{\bullet}/M_{bulge} \sim 0.002$ with a scatter of $\sim0.3~dex$.

The $M_{\bullet} - M_{bulge}$ relationship has been investigated at higher redshifts up to $z\sim4$. For example, \citet{Mclure2006} investigated the relationship in radio-loud galaxies at $z\sim 2$, \citet{Decarli2010b} studied a sample of 96 quasars out to $z\sim3$, \citet{Peng2006a,Peng2006b} used both gravitationally lensed and non-lensed galaxies to study the relationship out to $z\sim4.5$, while \citet{Targett2012} studied $z\sim4$ quasars. The high redshift studies all agree that the $M_{\bullet}/M_{bulge}$ relationship appears to be evolving with redshift. This evolution was seen to be independent of radio-loudness and quasar luminosity \citep[see][who studied both radio-loud and radio-quiet quasars, as well as investigated possible biases]{Decarli2010b}. The results of these studies agree with one another and imply that the black holes at high redshifts are more massive for a given bulge mass than their local counterparts. 

Conversely, the opposite was found in $z\sim2$ SMGs, \citep{Alexander2008}, where it was found that the SMGs lie below the local $M_{\bullet}/M_{bulge}$ relationship. This illustrates that the selection of objects has significant effects as to where they will be placed with respect to the local relationship. SMGs are biased towards extreme star-formation rates, which indirectly translates to a bias towards relatively massive galaxies. On the other hand, selection of quasars and radio galaxies is biased towards massive black holes. It is therefore perhaps unsurprising that these selected populations are biased towards different sides of the local $M_{\bullet}/M_{bulge}$ relationship.
 
While our results have considerable uncertainty, they are consistent (within the scatter) with what these other groups have found \citep{Peng2006b,Mclure2006,Decarli2010b,Targett2012}. For AMS12 to evolve to the local relationship, the bulge would have to grow $\sim3$ times as much as the central black hole from $z=2.8$ to $z=0$. 

Given the amount of molecular gas implied by the CO observations, if AMS12 were to evolve secularly, it would, at most, only increase its bulge mass by $\sim15\%$. Mergers could add more gas for star formation while adding stellar and black hole mass. It is expected that massive galaxies ($M_{\star}\grtsim10^{11}$\msol), undergo $>1$ mergers from $z\sim3$ to present day \citep{Conselice2003,Conselice2007,Bluck2009,Hopkins2010,Robaina2010}. However, it is not clear how mergers affect the black hole mass and whether it is possible to achieve the necessary growth of the bulge relative to the growth of the black hole. 

Alternatively, \citet{Decarli2010b} addressed the possibility that the remnants of the high redshift quasars they studied are high-mass outliers to the local relationship. These high redshift quasars are progenitors to present day massive ellipticals, and keeping their $M_{\bullet}/M_{bulge}$ value to $z=0$, they become outliers, rather than evolve, to the local relationship.

We must note that there was significant bias towards selecting a powerful quasar, giving a large $M_{\bullet}$, while demanding a faint 3.6 $\mu m$ flux, limiting the host galaxy's luminosity \citep{Martinez2005, Martinez2006b}. In addition, the search for CO in this object was initiated by selecting the brightest MAMBO detection from the obscured quasar sample \citep{Martinez2009}. 

There is also a possibility that by biasing ourselves towards such a high \lbol, AMS12 is super-Eddington, in which case we would be overestimating the black hole mass. If AMS12 were accreting at super-Eddington rates, the black hole mass would be overestimated by the amount by which the bolometric luminosity exceeds the Eddington-limited luminosity.  

\subsection{Comparison to other galaxies}
AMS12 is the first unlensed, high redshift source detected in both [CI] lines. This eliminates any ambiguity on the effects of possible differential magnification from lensing. 

The characterisation of the gas and dust in AMS12 is consistent with the observations of other high redshift galaxies, including the value of log($L_{\rm{FIR}}/\rm{L^{T}_{CO}}$), \citep[known as the star formation efficiency, i.e. Figure 8 in][]{SolVan2005}, where AMS12 lies within the scatter of the high redshift galaxies. The gas masses in these sources are similar to AMS12 \citep[i.e. studies of SMGs, ULIRGs and quasars at high redshifts][]{SolVan2005,Greve2005,Riechers2006,Coppin2008,Yan2010,Lacy2011}. 

A study of CO(1-0) in two $z\sim2.8$ obscured quasar hosts found similar gas and dust masses to AMS12 \citep{Lacy2011}. The studies of obscured quasars and their hosts also indicate mature systems, with dust and gas masses low compared to the stellar mass estimates \citep[e.g.][]{Lacy2011}. 

However, comparing to other high redshift quasars which are strongly lensed, is difficult as is illustrated with F10214. The gas and dust properties of F10214 are very similar to AMS12, with approximately equivalent [CI] abundances, cooling rate, and line ratios in terms of both CO and [CI]. 

\citet{Ao2008} modelled the CO emission in F10214 with LVG models, and found a similar dust temperature to AMS12 (80 K), and determined a range of gas kinetic temperatures from 45-80 K (note their Figure 7 has a similar shape to the high temperature region in Figure \ref{fig:cont_plots}), with $T_{\rm{ex}}$ of [CI] $\sim42$ K. 

\citet{Riechers2011a} detected CO(1-0) in F10214, and found there was no evidence for an extended, low-excitation gas component. For their analysis they assumed a constant magnification factor for the CO(1-0) (the magnification given by the higher-$J$ CO lines), hence if there is differential magnification of the gas components, their results may be affected. 

Deane et al. (2012a, in prep.), have revised the lens model in this F10214 and have indeed found differential magnification on frequency and spatial scales. Deane et al. (2012b, in prep.), study resolved CO(1-0) in this object and find preferential magnification between individual channels and predict distortion of the CO SED. Thus, AMS12 which is unlensed, offers, so far, a unique opportunity to study the gas and dust in an obscured quasar host without the added complication of gravitational lensing.  

\section{Summary}

In this paper we have presented new observations of the obscured quasar AMS12 and, along with previous mm and submm observations, we have investigated the dust and gas properties of this object. 

The FIR dust observations are well fit by a single component graybody model with dust temperature T$_{\rm{D}}$ = $88\pm8$ K, emissivity index $\beta$ = $0.6\pm0.1$, and \lfir\ = $(3.2\pm0.7)\times10^{13}$ \lsol, implying heating by the AGN.

The CO SED was fit with LVG models, and we used the marginalised PDF of the dust temperature as the prior distribution of the gas kinetic temperature to constrain the parameters. This yielded the gas kinetic temperature of T$_{\rm{G}}$ = \margpriorT\ K, and density n(H$_{2}$) = \margpriorN\ cm$^{-3}$, suggesting that SF is not the sole heating source.  

The atomic carbon fine structure lines \lowCI\ and \highCI, were observed and the [CI] excitation temperature was determined to be $43\pm10$ K, which is significantly lower than T$_{\rm{G}}$, indicating either [CI] is not in LTE, or it is from a more extended, lower temperature gas component.

The gas mass found from the CO(1-0) estimate, to be $\sim4.9\times10^{10}$ \msol. The dynamical mass was calculated from the CO linewidth to be $M_{\rm{dyn}}\rm{sin}^{2}\textit{i} =(2.7\pm0.2)\times10^{9}$ \msol\ assuming $r=0.8$ kpc, giving a limit to the host galaxy's inclination $\textit{i}\lesssim13 \degree$.

The stellar mass in this object is estimated at $M_{\star}=(3.2\pm0.3)\times10^{11}$ \msol. It follows that; the gas and dust mass are only a fraction of the current stellar mass. The $M_{\bullet}/M_{\star}$ ratio is $\grtsim0.005$, higher than in the local Universe. 

The system has already amassed the majority of its stellar mass and is host to a massive black hole, indicating a mature system. It is not clear how the system will evolve to the present-day $M_{\bullet}/M_{\star}$ relation, or whether the extreme value is due to a selection bias.

\bigskip

We gladly thank the staff at IRAM in particular Sascha Trippe and Chin-Shin Chang, for their assistance in the observations and reductions, and Christian Henkel for providing his LVG models for use in this paper. We thank Roger Deane for valuable insight into lensing and its effects, and Alex Karim for his helpful comments. We thank the anonymous referee for their useful comments and suggestions. H.S. and A.M.-S. are funded by SEPnet. A.M.-S. gratefully acknowledges a Post-Doctoral Fellowship from the United Kingdom Science and Technology Facilities Council, reference ST/G004420/1. This work is based in part on observations made with the Spitzer Space Telescope, which is operated by the Jet Propulsion Laboratory, California Institute of Technology under a contract with NASA.

\bibliographystyle{mn2e}

\label{lastpage} 

\end{document}